\documentclass[11pt,a4paper,preprintnumbers,nofootinbib]{revtex4}
\usepackage[T1]{fontenc}
\usepackage[utf8]{inputenc}
\usepackage{mathtools}
\usepackage{amsmath}
\usepackage{amsthm}
\usepackage{amsmath}
\usepackage{amssymb}
\usepackage{amstext}
\usepackage[final]{graphicx}
\usepackage{float}
\usepackage{booktabs}
\usepackage{fixltx2e}
\usepackage{subfigure}
\usepackage{siunitx}

\usepackage{array}
\newcolumntype{C}[1]{>{\centering\arraybackslash$}p{#1}<{$}}

\usepackage[dvipsnames]{xcolor}
\usepackage{slashed}
\usepackage[section]{placeins}
\usepackage{tabularx}
\PassOptionsToPackage{hyphens}{url}
\usepackage{hyperref}
\usepackage{makecell}	
\allowdisplaybreaks
\usepackage[normalem]{ulem}

\newcommand{\MG}{\textsc{MadGraph5}\_aMC@NLO\ } 
\definecolor{lightgray}{gray}{0.91}

\begin{document}
\preprint{DO-TH 20/14}

\title{Top and Beauty synergies in SMEFT-fits at present and future colliders }
\author{Stefan~Bi{\ss}mann}
\author{Cornelius~Grunwald}
\author{Gudrun~Hiller}
\author{Kevin~Kr{\"o}ninger}
\affiliation{Fakult\"at Physik, TU Dortmund, Otto-Hahn-Str.4, D-44221 Dortmund, Germany}

\begin{abstract}
	We perform  global fits within Standard Model Effective Field Theory (SMEFT) combining top-quark pair production processes and decay
	with  $b\rightarrow s$ flavor changing neutral current transitions and $Z \to b \bar b$ in three stages: using existing data from the LHC and $B$--factories, using projections for the HL-LHC and Belle II, and studying the additional new physics  impact from
	 a future  lepton collider. The latter is ideally suited to directly probe $\ell^+\ell^-\rightarrow t\bar t$ transitions.
	We observe powerful synergies in combining both  top and beauty observables
	as flat directions are removed and more operators can be probed.
    We find that a future  lepton collider significantly  enhances this interplay and qualitatively improves global SMEFT fits.
\end{abstract}

\maketitle

%======================================================================================================================================================
%

\section{Introduction}

Physics beyond the Standard Model (BSM) has and is being intensively searched for  at the Large Hadron Collider (LHC) and predecessor machines. However, despite the large amount of data analyzed, no direct detection of BSM particles has been reported to date.
Thus, BSM physics could be feebly interacting only, has signatures not covered by the standard searches, or is simply sufficiently separated from the electroweak scale.
The latter scenario opens up a complementary approach to  hunt for BSM  physics at high energy colliders, in a similar spirit as the high luminosity flavor physics programs
in pursuit of the precision frontiers with  indirect searches.
In this regard, the Standard Model Effective Field Theory (SMEFT) offers both a systematic and model-independent way to parametrize BSM contributions 
in terms of higher-dimensional  operators constructed out of Standard Model (SM) fields and consistent with SM symmetries. 
At energies below the scale of BSM physics, $\Lambda$, this framework allows to perform global fits which could give hints for signatures of BSM physics 
in different observables and sectors  simultaneously.

In recent years, this approach played a crucial role in the top-quark sector of SMEFT  \cite{Degrande:2018fog,Chala:2018agk,Durieux:2014xla,AguilarSaavedra:2010zi,DHondt:2018cww,Durieux:2018ggn,Buckley:2015nca,Buckley:2015lku,deBeurs:2018pvs,Brown:2019pzx,AguilarSaavedra:2018nen,Hartland:2019bjb,Maltoni:2019aot,Durieux:2019rbz,Neumann:2019kvk, Brivio:2019ius,Dror:2015nkp}.  
The SMEFT framework also allows the combination of top-quark data with $B$ data \cite{Bissmann:2019gfc, Aoude:2020dwv, Fox:2007in,Grzadkowski:2008mf,Drobnak:2011aa,Brod:2014hsa},
which, thanks to different sensitivities, significantly improves constraints on SMEFT coefficients  \cite{Bissmann:2019gfc}.

In this work, we extend previous works and analyze sensitivities to semileptonic four-fermion operators. The reason for doing so  goes
way beyond of making the fit more model-independent:
Firstly, semileptonic four-fermion operators are presently of high  interest as they are the agents of the flavor anomalies, hints of a breakdown of the SM in semileptonic $b \to s$ decay data \cite{Bifani:2018zmi}.
Secondly, these operators provide contact interactions of top quarks and leptons, which could be studied ideally at future lepton colliders, {\it e.g.} ILC~\cite{Amjad:2015mma, Amjad:2013tlv}, CLIC~\cite{Abramowicz:2018rjq} or FCC~\cite{Abada:2019zxq}, as discussed in Refs.~\cite{Kane:1991bg, Atwood:1991ka, Grzadkowski:1997cj, Brzezinski:1997av, Boos:1999ca, Jezabek:2000gr, Grzadkowski:2000nx, Janot:2015yza, Rontsch:2015una, Khiem:2015ofa, Englert:2017dev, Durieux:2018tev, Cao:2015qta}. 
We  intend to specifically work out and detail  the interplay of constraints for  operators with gauge bosons, that is, covariant derivatives in the SMEFT language,
and four fermion operators  in top-pair production processes, $Z \to b \bar b$  and $b \to s$ transitions for
three stages: today, combining existing LHC, $Zbb$ and $B$-factory data,  near future, adding  projections from HL-LHC \cite{Atlas:2019qfx} and Belle II \cite{Kou:2018nap}, and the far future, putting all together with lepton collider input,  for the concrete example of  CLIC~\cite{Abramowicz:2018rjq}; we investigate how  a future lepton collider impacts  constraints and 
opens up new directions for testing BSM physics precisely.

This work is organized as follows:
In Sec.~\ref{sec:EFTs} we introduce the dimension-six SMEFT operators considered in this work and the low-energy effective field theories (EFTs) employed to compute SM and BSM contributions to $B$ observables.
We also present the matching between SMEFT and weak effective theory (WET) and highlight how $SU(2)_L$ invariance of the SMEFT Lagrangian allows to relate top-quark physics and $b\rightarrow s$ flavor-changing neutral currents (FCNCs).
In Sec.~\ref{sec:Sensitivity} we discuss the sensitivity of different observables to the various effective operators considered.
Fits to present top-quark, $Zbb$, and $B$ data are presented in Sec.~\ref{sec:fit-now}. 
We analyze how the complementary sensitivity of the observables from top-quark, $Zbb$, and $B$ sectors improves constraints on the SMEFT coefficients.
In Sec.~\ref{sec:fit-future} we consider different future scenarios, and detail on the question how measurements at a future lepton collider can provide additional information on SMEFT coefficients. 
In Sec.~\ref{sec:conclusion} we conclude. Additional information is provided in several appendices.

%============================================================================================================================================================
\section{ Effective theory setup}
\label{sec:EFTs}

In this section  we give the requisite  EFT setup to describe BSM contributions to top-quark and beauty observables.
We introduce the SMEFT Lagrangian in Sec.~\ref{sec:SMEFT},
and identify the effective operators contributing to interactions of third-generation quarks.
Consequences for FCNCs that arise from flavor mixing are worked out in Sec.~\ref{sec:SMEFT-SSB}, where
we also  highlight the complementarity between contributions from up-type and down-type quarks.
The matching conditions for  $B$ observables in the low energy effective Lagrangian in terms of SMEFT coefficients  are detailed in Sec.~\ref{sec:Match}.

%============================================================================================================================================================

\subsection{SMEFT dimension-six operators}
\label{sec:SMEFT}

At energies sufficiently below the scale of new physics, $\Lambda$, the effects of new interactions and BSM particles can be described by a series of higher-dimensional effective operators with mass dimension $d>4$ \cite{Weinberg:1978kz, Buchmuller:1985jz}. 
These operators are built out of SM fields and respect the symmetries of the SM. 
The SMEFT Lagrangian $\mathcal{L}_\text{SMEFT}$ is obtained by adding these  $d$-dimensional operators $O_i^{(d)}$ together 
 with corresponding Wilson coefficients $C_i^{(d)}$ to the SM Lagrangian $\mathcal{L}_\text{SM}$. The $C_i^{(d)}$ encode the BSM couplings and, 
in order to be dimensionless, require a factor of  $\Lambda^{4-d}$.
The leading SMEFT contributions arise at dimension six:
\begin{align}
    \mathcal{L}_\textmd{SMEFT}=\mathcal{L}_\textmd{SM}+\sum_i\frac{C^{(6)}_i}{\Lambda^2}O_i^{(6)}+\mathcal{O}\left(\Lambda^{-4}\right)\,.
    \label{eq:L_SMEFT}
\end{align}
Contributions from odd-dimensional operators lead to lepton- and baryon-number violation \cite{Degrande:2012wf, Kobach:2016ami} and are neglected in 
this work.
In the following, we employ the \textit{Warsaw} basis \cite{Grzadkowski:2010es} of dimension-six operators, and consider operators  with gauge bosons
\begin{align}
    \begin{aligned}
        &O_{\varphi q}^{(1)}=\left(\varphi^\dagger i\overleftrightarrow D_\mu \varphi\right)\left(\bar q_L \gamma^\mu q_L\right)\,,\quad
        O_{\varphi q}^{(3)}=\left(\varphi^\dagger i\overleftrightarrow D_\mu^I \varphi\right)\left(\bar q_L \tau^I\gamma^\mu q_L\right)\,, \\
        &O_{uB}=\left(\bar{q}_L\sigma^{\mu\nu}u_R\right)\tilde{\varphi}B_{\mu\nu}	\,,
        \quad O_{uW}=\left(\bar{q}_L\sigma^{\mu\nu}\tau^{I}u_R\right)\tilde{\varphi}W_{\mu\nu}^{I}	\,,
        \quad O_{uG}=\left(\bar{q}_L\sigma^{\mu\nu}T^{A}u_R\right)\tilde{\varphi}G_{\mu\nu}^{A}	\,,\\
        &O_{\varphi u}=\left(\varphi^\dagger i\overleftrightarrow D_\mu \varphi\right)\left(\bar u_R \gamma^\mu u_R\right)\,,\quad
        \label{eq:top_boson}
    \end{aligned}
\end{align}
and semileptonic four-fermion operators 
\begin{align}
    \begin{aligned}
        &O_{lq}^{(1)}=\left(\bar l_L \gamma_\mu l_L\right)\left(\bar q_L \gamma^\mu q_L\right)	\,,\quad
        O_{lq}^{(3)}=\left(\bar l_L \gamma_\mu\tau^I l_L\right)\left(\bar q_L \gamma^\mu\tau^I q_L\right)	\,,\quad
        O_{qe}=\left(\bar q_L \gamma_\mu q_L\right)\left(\bar e_R \gamma^\mu e_R\right)	\,,\\
        &O_{eu}=\left(\bar e_R \gamma_\mu e_R\right)\left(\bar u_R \gamma^\mu u_R\right)	\,,\quad
        O_{lu}=\left(\bar l_L \gamma_\mu l_L\right)\left(\bar u_R \gamma^\mu u_R\right)	\,.
        \label{eq:top_fermion}
    \end{aligned}
\end{align}
Here, 
$q_L$, $l_L$ are the quark and lepton  $SU(2)_L$ doublets, and $u_R$, $e_R$  the up-type quark and charged lepton $SU(2)_L$ singlets, respectively.
Flavor indices  that exist for each SM fermion field are suppressed here for brevity but will be discussed in Sec.~\ref{sec:SMEFT-SSB}.
With $B_{\mu\nu}$, $W^I_{\mu\nu}$ and $G^A_{\mu\nu}$ we denote the gauge field strength tensors of $U(1)_Y$, $SU(2)_L$
and $SU(3)_C$, respectively. 
$T^A=\lambda^A/2$ and $\tau^I/2$ are the generators of $SU(3)_C$ and $SU(2)_L$ in the fundamental representation with $A=1,\dots,8$ and $I=1,2,3$, 
and $\lambda^A$ and $\tau^I$ are the Gell--Mann and Pauli matrices, respectively.
The SM Higgs doublet  is denoted by $\varphi$ with its conjugate  given as $\tilde\varphi= i \tau^2 \varphi$,
$ \left(\varphi^\dagger i\overleftrightarrow D_\mu \varphi\right)=i\varphi^\dagger( D_\mu \varphi )-  i( D_\mu \varphi^\dagger) \varphi $ and $\left(\varphi^\dagger i\overleftrightarrow D_\mu^I \varphi\right)=i\varphi^\dagger \tau^I ( D_\mu \varphi )-  i( D_\mu \varphi^\dagger) \tau^I \varphi $.

Further dimension-six operators exist that contribute at subleading order to top-quark observables  such as 
dipole operators $O_{dX}$ with $X=B,W,G$ and right-handed $b$ quarks, with  contributions  suppressed by $m_b/m_t$.
We neglect those as well as all other SMEFT operators involving right-handed down-type quarks.
Scalar and tensor operators  $O^{(1/3)}_{lequ}$ are not included in our analysis since these operators do not give any relevant contributions at $\mathcal{O}(\Lambda^{-2})$ for the interactions considered in this work \cite{Durieux:2019rbz, Durieux:2018tev}.
Contributions from four-quark operators to $t\bar t\gamma$, $t\bar t Z$ and $t\bar t$ production are neglected as $t\bar t$ production at the LHC is dominated by the 
$gg$ channel \cite{Buckley:2015lku}~\footnote{Differential cross sections on the other hand are sensitive to four-fermion contributions \cite{Brivio:2019ius}. Since
bin-to-bin correlations are not available, yet important~\cite{Bissmann:2019qcd},  we do not consider such observables in our fit.}.
In addition we also neglect leptonic dipole operators, {\it  i.e.}, vertex corrections to lepton currents because they are severely constrained by $Z$-precision measurements \cite{Zyla:2020zbs}.

Note that dipole operators are in general non-hermitian which allows for complex-valued Wilson coefficients. 
However, the dominant interference terms are proportional only to the real part of the coefficients. 
For the sake of simplicity, we thus assume all coefficients to be real-valued. 

\subsection{Flavor and mass basis}
\label{sec:SMEFT-SSB}

The dimension-six operators (\ref{eq:top_boson}), (\ref{eq:top_fermion}) are given in the flavor basis. 
In general, quark mass and flavor bases are related by unitary transformations $S^k_{L/R}$, $k=u,d$,
\begin{align}
    u_{L/R}^i = \left(S^u_{L/R}\right)_{ij} u^{\prime j}_{L/R}\,,\quad 
    d_{L/R}^i = \left(S^d_{L/R}\right)_{ij} d^{\prime j}_{L/R}\,,
\end{align}
where $u'$ and $d'$ denote up- and down-type quarks in the mass basis, respectively, and $i,j=1,2,3$ are flavor indices. The CKM matrix  $V$ is then given  as
\begin{align}
    V = \left(S^u_L\right)^\dagger S^d_L\,.    
\end{align}
The rotation matrices of right handed quarks $S_R^{u/d}$ can simply be absorbed in the flavor-basis Wilson coefficient $C_i$, 
giving rise to coefficients in the mass basis, denoted by $\hat C_i$  \cite{Aebischer:2015fzz}.
In contrast, the flavor rotations $S^{u/d}_L$ of quark doublets relate different physical processes by $SU(2)_L$-symmetry.
Consider a contribution involving a doublet quark current with $SU(2)_L$-singlet structure, {\it i.e.}, the $C^{(1)} O^{(1)}$ terms with quark flavor indices restored.
For instance, 
\begin{align}
    \begin{aligned}
    C_{\varphi q}^{(1)ij} O_{\varphi q}^{(1)ij} &= C_{\varphi q}^{(1)ij} \left(\varphi^\dagger i\overleftrightarrow D_\mu \varphi\right)\left(\bar u_L^i \gamma^\mu u_L^j + \bar d_L^i \gamma^\mu d_L^j\right)\\
    &=C_{\varphi q}^{(1)ij} \left(\varphi^\dagger i\overleftrightarrow D_\mu \varphi\right)\left(\left(S^{u\dagger}_L\right)_{ki}\bar u_L^{\prime k} \gamma^\mu \left(S^{\vphantom{\dagger}u}_L\right)_{jl} u_L^{\prime l} + \left(S^{d\dagger}_L\right)_{mi}\bar d_L^{\prime m} \gamma^\mu \left(S^{\vphantom{\dagger}d}_L\right)_{jn} d_L^{\prime n}\right)\\
    &=\hat C_{\varphi q}^{(1)kl} \left(\varphi^\dagger i\overleftrightarrow D_\mu \varphi\right)\left(\bar u_L^{\prime k} \gamma^\mu  u_L^{\prime l} + V^\dagger_{mk} V^{\vphantom{\dagger}}_{ln} \bar d_L^{\prime m} \gamma^\mu  d_L^{\prime n}\right)\,.
    \label{eq:mass-operator-relation}
    \end{aligned}
\end{align}
Since we are interested in top-quark physics, in the last line we  have chosen to work in the up-mass basis, the basis in which up-quark flavor and mass bases are 
identical and flavor mixing is entirely in the down-sector. Irrespective of this choice for the mass basis,
 $ C^{(1)ij}_{\varphi q}$ induces in general contributions to both $u^i-u^j$ and $d^i-d^j$ transitions.
In the up mass basis, $d^i-d^j$ transitions come with additional CKM-matrix elements.
Contributions involving a doublet quark current with $SU(2)_L$-triplet structure, {\it i.e.} the $C^{(3)} O^{(3)}$ terms
have an additional minus sign between the up-sector and down-sector currents,
\begin{align}
    \begin{aligned}
    C_{\varphi q}^{(3)ij} O_{\varphi q}^{(3)ij} &= \hat C_{\varphi q}^{(3)kl} \left(\varphi^\dagger i\overleftrightarrow D_\mu^3 \varphi\right)\left(\bar u_L^{\prime k} \gamma^\mu  u_L^{\prime l} - V^\dagger_{mk} V^{\vphantom{\dagger}}_{ln} \bar d_L^{\prime m} \gamma^\mu  d_L^{\prime n}\right)\,.
    \label{eq:mass-operator-relation3}
    \end{aligned}
\end{align}
As a result, up-type and down-type quarks probe different combinations of $C^{(1)}$ and $C^{(3)}$, a feature recently also exploited in probing lepton flavor universality and conservation with processes involving neutrinos \cite{Bause:2020auq}.
Further details on SMEFT coefficients and operators in the up-mass basis are given in App.~\ref{app:mass-basis} and App.~\ref{app:relations}, respectively.

In this analysis, we only consider contributions from (flavor basis) Wilson coefficients with third generation quarks, $\hat C_i^{33}$.
Such hierarchies may arise in BSM scenarios with enhanced couplings to third-generation quarks, similar to the top-philic scenario discussed in Ref.~\cite{AguilarSaavedra:2018nen}.
As can be seen in Eqs.~\eqref{eq:mass-operator-relation}, \eqref{eq:mass-operator-relation3}, 
flavor mixing induces contributions to $d_L^i \rightarrow d_L^j$ transitions for  $i \neq j$ with CKM suppressions $V_{ti} V_{tj}^*$, just like the SM.
In this work, we include FCNC data from $b\rightarrow s$ transitions, while $s\to d$ transitions do presently not yield more significant constraints \cite{Aoude:2020dwv}, and are not considered further.
This leaves us with eleven real-valued SMEFT coefficients for the global fits
\begin{align}
	\begin{aligned}
		\hat C^{33}_{uB}\,,\enskip
		\hat C^{33}_{uG}\,,\enskip
		\hat C^{33}_{uW}\,,\enskip 
		\hat C^{(1)33}_{\varphi q}\,,\enskip
		\hat C^{(3)33}_{\varphi q}\,,\enskip
		\hat C^{33}_{\varphi u}\,,\enskip
		\hat C^{33}_{eu}\,,\enskip
        \hat C^{33}_{lu}\,,\enskip
        \hat C^{33}_{qe}\,,\enskip
        \hat C^{(1)33}_{lq}\,,\enskip 
		\hat C^{(3)33}_{lq}\,,
		\label{eq:future-SMEFT-coff}
	\end{aligned}
\end{align}
defined in the up-mass basis. 

Lepton universality does not have to be  assumed for fits  to present data 
since the bulk of the existing $b$--physics precision distributions is with muons. In the future, Belle II is expected to
deliver both $b \to s e^+ e^-$ and $b \to s \mu^+ \mu^-$ distributions, and to shed light on the present hints that electrons and muons 
may be more different than thought \cite{Hiller:2014qzg}. 
In the far future,
the $b \to s e^+ e^-$  results can  be combined with $t \bar t$-production data  from an $e^+ e^-$-collider;
the muon ones could be combined with data from a muon collider, to improve the prospects for lepton flavor-specific fits.
We also note that lepton flavor violating operators could also be included in the future.
On the other hand,  once data on  dineutrino modes are included in the fit,
assumptions on lepton flavor are in order, since the branching ratios are measured in a flavor-inclusive way:
\begin{align}
{\cal{B}}(b \to s \nu \bar \nu)=\sum_{i,j} {\cal{B}}(b \to s \nu_i \bar \nu_j) \, .
\end{align}
Universality dictates that the total dineutrino branching ratio is given by three times a flavor-specific one,
${\cal{B}}(b \to s \nu \bar \nu)=3  {\cal{B}}(b \to s \nu_i \bar \nu_i)$. Here, $i$ is fixed, but could be any of the three flavors.
We do assume universality when we include dineutrino modes in the fits to future data.

\begin{figure}[t]
    \centering
    \includegraphics{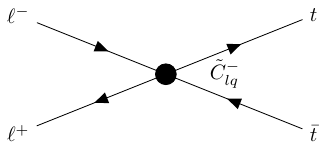}\hspace{0.2cm}
    \includegraphics{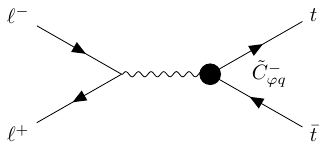}\\
    \vspace{0.4cm}
    \includegraphics{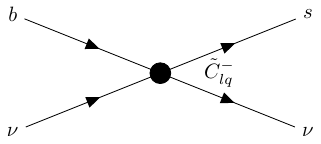}\hspace{0.2cm}
    \includegraphics{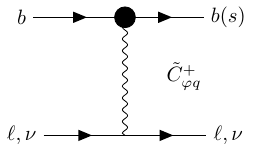}\hspace{0.2cm}
    \includegraphics{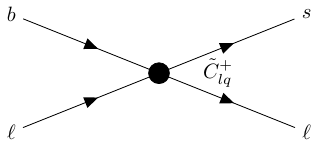}
    \caption{Sensitivities to $ \tilde C_{lq}^\pm$ and $ \tilde C_{\varphi q}^\pm $, defined in (\ref{eq:def-Cpm}),  in top-quarks with charged leptons (upper row),
    and beauty with charged leptons and neutrinos (lower row). The black circles denote SMEFT operators, wavy lines are electroweak gauge bosons.
    }
        \label{fig:comp}
\end{figure}

As is customary, in the following  we use rescaled coefficients and drop the superscript for brevity
\begin{align}
    \tilde C_i = \frac{v^2}{\Lambda^2} \hat C_i^{33}\,,
\end{align}
where $v=\SI{246}{\GeV}$ is the Higgs vacuum expectation value.
To highlight $SU(2)_L$ complementary between top and beauty, we introduce 
\begin{align}
    \tilde C_{lq}^\pm = \tilde C^{(1)}_{lq} \pm  \tilde C^{(3)}_{lq}\,,\quad 
    \tilde C_{\varphi q}^\pm = \tilde C^{(1)}_{\varphi q} \pm  \tilde C^{(3)}_{\varphi q}\,.
    \label{eq:def-Cpm}
\end{align}
The sensitivities are illustrated  in Fig.~\ref{fig:comp}.

%=============================================================================

\subsection{Matching and Running: SMEFT and WET}
\label{sec:Match}

To constrain the Wilson coefficients of the SMEFT operators in Eqs.~\eqref{eq:top_boson} and \eqref{eq:top_fermion} using $B$ physics measurements, the SMEFT Lagrangian has to be matched onto the WET Lagrangian, see  App.~\ref{sec:WET} for details. The procedure to compute BSM contributions at the scale $\mu_b$ in terms of coefficients given at a higher scale $\mu$ is described in detail in Ref.~\cite{Bissmann:2019gfc} and adapted here.
Throughout this work, we consider values for Wilson coefficients at the scale $\mu=1\,\textmd{TeV}$.

\subsubsection{SMEFT RGE}
The values of the Wilson coefficients depend on the energy scale $\mu$ of the process considered.
The renormalization group equations (RGEs) allow to combine measurements at different scales in one analysis.
The RGEs for  Eqs.~\eqref{eq:top_boson} and \eqref{eq:top_fermion} have been computed in Refs.~\cite{Jenkins:2013sda, Jenkins:2013zja, Jenkins:2013wua, Alonso:2013hga}. 
We include these effects at one-loop level by applying the \texttt{wilson} \cite{Aebischer:2018bkb} package.
%We consider the leading contributions at $\mathcal{O}(\alpha_s)$. 
%For the dipole operators $O_{uW}$, $O_{uB}$ and $O_{uG}$  the procedure is discussed in detail in Ref.~\cite{Bissmann:2019gfc}.  
%The other operators considered in this analysis do not run at $\mathcal{O}(\alpha_s)$. 

\subsubsection{Matching SMEFT onto WET}

Flavor rotations allow for contributions from $\hat C_i^{33}$ coefficients to $b\to s$ transitions whenever two $SU(2)_L$ quark doublets are present in the operator.
We obtain finite tree level contributions from $O^{(1)}_{\varphi q}$, $O^{(3)}_{\varphi q}$, $O^{(1)}_{lq}$, $O^{(3)}_{lq}$ and $O_{qe}$ to the WET coefficients
of the semileptonic four-fermion operators $O_{9,10,L}$, defined in  App.~\ref{sec:WET}, as \cite{Aebischer:2015fzz, Buras:2014fpa}:
\begin{align}
    \begin{aligned}
        \Delta C_9^\textmd{tree} &= \frac{\pi}{\alpha} \left[ \tilde C^{(1)}_{l q} + \tilde C^{(3)}_{l q} + \tilde C_{q e} + \left(-1+4\sin^2\theta_w\right)\left(\tilde C^{(1)}_{\varphi q} + \tilde C^{(3)}_{\varphi q}\right)\right] \simeq \frac{\pi}{\alpha} \left[   \tilde C_{lq}^+   +\tilde C_{q e} \right] \,,\\
        \Delta C_{10}^\textmd{tree} &= \frac{\pi}{\alpha} \left[- \tilde C^{(1)}_{l q} - \tilde C^{(3)}_{l q} + \tilde C_{q e} + \tilde C^{(1)}_{\varphi q} + \tilde C^{(3)}_{\varphi q} \right]  =   \frac{\pi}{\alpha} \left[-  \tilde C_{lq}^+  + \tilde C_{\varphi q}^+  +\tilde C_{q e} \right]   \,,\\
        \Delta C_{L}^\textmd{tree} &= \frac{\pi}{\alpha} \left[ \tilde C^{(1)}_{l q} - \tilde C^{(3)}_{l q} + \tilde C^{(1)}_{\varphi q} + \tilde C^{(3)}_{\varphi q}\right] =\frac{\pi}{\alpha} \left[  \tilde C_{lq}^- + \tilde C_{\varphi q}^+ \right]\,,
        \label{eq:match-tree}
    \end{aligned}
\end{align}
where $\sin^2 \theta_w =0.223$  denotes the weak mixing angle.
We used for $\Delta C_9^\textmd{tree}$  in the second step the well-know suppression of $Z$-penguins due to the numerical smallness of the $Z$'s vector coupling to charged leptons \cite{Buchalla:2000sk}.

In addition to these dominant contributions, SMEFT operators induce contributions to WET dipole operators $O_{7,8}$, semileptonic operators $O_{9,10,L}$ and 
$|\Delta B|=2$ mixing at one-loop level \cite{Aebischer:2015fzz,Bobeth:2017xry,Dekens:2019ept,Endo:2020kie, Hurth:2019ula, Aoude:2020dwv}:
\begin{align}
    &\begin{aligned}
        \Delta C_7^{\textmd{loop}}=&\frac{\sqrt{2}m_t}{m_W}\left[\tilde C_{uW}E_{7}^{uW}(x_t)+\tilde C_{uW}^*F_{7}^{uW}(x_t)
        +\frac{\cos\theta_w}{\sin\theta_w}\left(\tilde C_{uB}E_{7}^{uB}(x_t)+\tilde C_{uB}^*F_{7}^{uB}(x_t)\right)\right]\\&+E_7^{\varphi q}(x_t)\tilde C_{\varphi q}^{(1)} +  E_7^{\varphi q (3)}(x_t)\tilde C_{\varphi q}^{(3)}
        \label{Eq:MatchC7}
    \end{aligned}\\
    &\begin{aligned}
        \Delta C_{8}^\textmd{loop}=&\frac{\sqrt{2}m_t}{m_W}\left[\tilde C_{uW}E_{8}^{uW}(x_t)+\tilde C_{uW}^*F_{8}^{uW}(x_t)
        -\frac{g}{g_s}\left(\tilde C_{uG}E_8^{uG}(x_t)+\tilde C_{uG}^*F_8^{uG}(x_t)\right)\right]\\&+E_8^{\varphi q}(x_t)\tilde C_{\varphi q}^{(1)}+E_8^{\varphi q(3)}(x_t)\tilde C_{\varphi q}^{(3)}\,,    
        \label{Eq:MatchC8}
    \end{aligned}\\
    &\begin{aligned}
        \Delta C_9^{\textmd{loop}} =&
        \sqrt{2}\frac{m_t}{m_W}\left[\left(\frac{Y_{uW}(x_t)}{\sin^2\theta_w}-Z_{uW}(x_t)\right)\text{Re}(\tilde C_{uW})-\frac{\cos\theta_w}{\sin\theta_w}Z_{uB}(x_t)\text{Re}(\tilde C_{uB})\right] \\
        &+ \frac{1}{\sin^2\theta_w}\left\{ 
        I_1(x_t)\left[ \tilde C_{eu} + \tilde C_{lu} +(-1+4\sin^2\theta_w) \tilde C_{\varphi u} \right]    
        + I_2(x_t)\left[ \tilde C_{qe} + \tilde C_{lq}^{(1)} \right] \right.\\  
        &+ \left. I^{lq}(x_t) \tilde C_{lq}^{(3)}
        +  \left[ (-1+4\sin^2\theta_w) I_2(x_t) \tilde C^{(1)}_{\varphi q} + I_1^{\varphi q}(x_t) \tilde C^{(3)}_{\varphi q} \right]
        \right\}
        \label{Eq:MatchC9}
    \end{aligned}\\
    &\begin{aligned}
        \Delta C_{10}^{\textmd{loop}} =&
        -\frac{\sqrt{2}}{\sin^2\theta_w}\frac{m_t}{m_W}Y_{uW}(x_t)\text{Re}(\tilde C_{uW}) \\
        &+ \frac{1}{\sin^2\theta_w}\left\{ 
        I_1(x_t)\left[ \tilde C_{eu} - \tilde C_{lu} +  \tilde C_{\varphi u} \right]    
        + I_2(x_t)\left[ \tilde C_{qe} - \tilde C_{lq}^{(1)} \right] \right.\\  
        & \left. -I^{lq}(x_t) \tilde C_{lq}^{(3)}
        + \left[ I_2(x_t)(x_t) \tilde C^{(1)}_{\varphi q} + I_{2}^{\varphi q}(x_t) \tilde C^{(3)}_{\varphi q} \right]
        \right\}
        \label{Eq:MatchC10}
    \end{aligned}\\
    &\begin{aligned}
        \Delta C_{L }^\textmd{loop}=&I_{uW}^{\nu}\text{Re}(\tilde C_{uW}) + I^{\nu (1)}_{\varphi q} \tilde C^{(1)}_{\varphi q} + I^{\nu (3)}_{\varphi q} \tilde C^{(3)}_{\varphi q}  + I^{\nu }_{l u} (\tilde C_{\varphi u} + \tilde C_{lu})+ I^{\nu (1)}_{l q} \tilde C^{(1)}_{lq} + I^{\nu (3)}_{l q} \tilde C^{(3)}_{lq} 
        \,,
        \label{Eq:MatchCL}
    \end{aligned}\\
    &\begin{aligned}
        \Delta C_{1,tt}^\textmd{mix, loop}=&+\sqrt{2}\frac{m_t}{m_W} \text{Re}(\tilde C_{uW}) \frac{9x_t}{4} \left(\frac{x_t+1}{(x_t-1)^2}-\frac{2x_t}{(x_t-1)^3} \log x_t\right) +4 S_0(x_t)\tilde C_{\varphi q}^{(3)} \\&+ H_1(x_t) \tilde C^{(1)}_{\varphi q} + H_2(x_t) \tilde C^{(3)}_{\varphi q} \,,
        \label{Eq:MatchC1}
    \end{aligned}
\end{align}
which are present also in absence of CKM-mixing, and
with $x_t=m_t^2/m_W^2$. 
Explicit expressions for the $x_t$-dependent functions can be found in Refs.~\cite{Aebischer:2015fzz,Bobeth:2017xry,Dekens:2019ept,Endo:2020kie,Hurth:2019ula, Aoude:2020dwv}. 
For completeness, we also give these functions in App.~\ref{App:match}.

Note that there is  sensitivity, although only at the one-loop level, to the semileptonic operators with up-type singlet quarks, $O_{eu}$ and $O_{lu}$.
The numerical values of the matching conditions at $\mu_W=m_W$ are computed with \texttt{wilson} following Ref.~\cite{Dekens:2019ept} and are provided in App.~\ref{app:numerics}.
In the actual analysis, RGE effects in SMEFT and WET are taken into account as well.

\subsubsection{WET RGE}
 
We employ  \texttt{flavio} \cite{Straub:2018kue} and \texttt{wilson} to compute the values of the SM and BSM contributions 
at the scale $\mu_b$.

%=============================================================================

\section{Observables}

\label{sec:Sensitivity}

In this section we give details on how  theory predictions and distributions  for top-observables (Sec.~\ref{sec:top-obs}), $Z\to b\bar b$ transitions (Sec.~\ref{sec:Z-obs}), and $B$- physics (Sec.~\ref{sec:B-obs}) 
are obtained, and discuss  the sensitivities of the observables to SMEFT coefficients (Sec.~\ref{sec:sens-obs}).

%=============================================================================

\subsection{Top-quark observables \label{sec:top-obs}}

We employ  the Monte Carlo (MC) generator \MG \cite{Alwall:2014hca} to compute the $t\bar t$, $t\bar t \gamma$ and $t\bar t Z$ production cross sections at the LHC and the $t\bar t$ production cross section and the forward-backward symmetry at CLIC in LO QCD.
The cross sections can be parametrized in terms of the Wilson coefficients as
\begin{align}
    \sigma = \sigma^\mathrm{SM} + \sum_i \tilde C_i\sigma_i^\text{int.} + \sum_{i \leq j} \tilde C_i \tilde C_j \sigma_{ij}^\text{BSM}\,, 
    \label{eq:interpol}
\end{align}
where $\sigma_i^\textmd{int.}$ and $\sigma_{ij}^\text{BSM}$ denote interference terms between SM and dimension-six operators and purely BSM terms, respectively. 
The forward-backward asymmetry is defined as 
\begin{align}
    A_\textmd{FB} = \frac{\sigma_\textmd{FB}}{\sigma}\,,\quad \sigma_\textmd{FB} = \int_{-1}^1 \textmd{d}\cos\theta \,\textmd{sign}(\cos\theta)\frac{d \sigma}{d \cos\theta}\,,
    \label{eq:AFB-interpol}
\end{align}
where $\theta$ denotes the angle between the three-momenta of the top quark and the positron in the center-of-mass frame. BSM contributions in both  numerator and denominator are parametrized according to Eq.~\eqref{eq:interpol}. 

To obtain  $\sigma_i^\textmd{int.}$ and $\sigma_{ij}^\text{BSM}$  we utilize the \texttt{dim6top\_LO} UFO model \cite{AguilarSaavedra:2018nen}.
For the computation of the fiducial cross sections of $t\bar t \gamma$ production we generate samples as a $2\rightarrow7$ process including BSM contributions in the top-quark decay. The fiducal acceptances are obtained by showering the events with \texttt{PYTHIA8} \cite{Sjostrand:2014zea} and performing an event selection at particle level with \texttt{MadAnalysis} \cite{Conte:2012fm, Conte:2014zja, Dumont:2014tja}. For the jet clustering we apply the anti-$k_t$ algorithm \cite{Cacciari:2008gp} with radius parameter $R=0.4$ using \texttt{FastJet} \cite{Cacciari:2011ma}. The computation is discussed in detail in  Ref.~\cite{Bissmann:2019gfc}. 

We compute the helicity fractions according to Ref.~\cite{Zhang:2010dr} with the difference that we also include quadratic contributions. In our analysis, we consider only $O_{uW}$ as only this operator gives contributions $\mathcal{O}(\Lambda^{-2})$ that are not suppressed by a factor $m_b/m_t$.
The top-quark decay width is computed following Ref.~\cite{Zhang:2014rja} including quadratic contributions. 

%=============================================================================

\subsection{\boldmath$Zb\bar b$ observables \label{sec:Z-obs}}

To compute $Z\rightarrow b\bar b$ observables we employ \MG together with the \texttt{dim6top\_LO} UFO model for both the forward-backward asymmetry $A^{0,b}_\textmd{FB}$ and the ratio of partial widths for $Z\rightarrow f\bar f$ 
\begin{align}
    R_b=\frac{\Gamma_{b\bar b}}{\Gamma_\textmd{had}}\,,\quad \Gamma = \Gamma^\textmd{SM} + \sum_i \tilde C_i \Gamma_i^\textmd{int} +\sum_{i \leq j} \tilde C_i \tilde C_j  \Gamma_{ij}^\textmd{BSM}\,.
\end{align}
BSM contributions to $A^{0,b}_\textmd{FB}$ are computed using Eq.~\eqref{eq:AFB-interpol}, and for $R_b$ we include BSM contributions in both  numerator and denominator.

%=============================================================================

\subsection{\boldmath$B$-physics observables \label{sec:B-obs}}

For observables in $b\rightarrow s\gamma$ and $b\rightarrow s\ell^+\ell^-$ transitions we employ \texttt{flavio} together with the \texttt{wilson} package to compute the BSM contributions in terms of $\Delta \bar C_i^{(0)} (\mu_W)$ at the scale $\mu_W = m_W$. For $b \to s \nu \bar \nu$ the Wilson coefficient $C_L$ does not run.
BSM contributions are considered at LO in $\alpha_s$ and run with \texttt{wilson} in the WET basis from the scale $\mu_W$ to $\mu_b$, at which the observables are computed. 
To compute the observables for different values of the SMEFT Wilson coefficients $\tilde C_i$, they are run from the scale $\mu$ to $\mu_W$ and matched onto the WET basis according to Eqs.~\eqref{eq:match-tree}-\eqref{Eq:MatchC10}. 

Branching ratios of $b\rightarrow s\nu\bar\nu$ transitions  are computed via \cite{Buras:2014fpa}
\begin{align}
    \text{BR}(B\rightarrow K^{(*)}\nu\bar\nu) =  \text{BR}(B\rightarrow K^{(*)}\nu\bar\nu)_\textmd{SM} \frac{\left|\Delta C_L(\mu_b) - C_L(\mu_b)_\textmd{SM} \right|}{C_L(\mu_b)_\textmd{SM}}\,,
\end{align}
where 
\begin{align}
    \begin{aligned}
    \text{BR}(B^+\rightarrow K^{+}\nu\bar\nu)_\textmd{SM} &= (4.0\pm0.5)\times 10^{-6}\,,\\
    \quad  \text{BR}(B^0\rightarrow K^{0 *}\nu\bar\nu)_\textmd{SM} &= (9.2\pm1.0)\times 10^{-6}\,,
    \end{aligned}
\end{align}
and $C_L(\mu_b)_\textmd{SM} =\frac{X_s}{\sin^2\theta_w}$ 
with $X_s = 1.469\pm 0.017$, and lepton flavor universality is assumed. 

We also consider the $B_s-\bar B_s$ mass difference $\Delta M_s$, which can be computed as \cite{DiLuzio:2019jyq}
\begin{align}
    \label{eq:DeltaMs}
    \Delta M_s = \Delta M_s^\textmd{SM} \left|1+\frac{\Delta C_{1,tt}^\text{mix} (\mu_W)}{S_0(x_t)}\right|\,,
\end{align}
where $S_0$  denotes  the Inami-Lim function. We employ the SM value  $\Delta M_s^\textmd{SM} = \left(18.4 ^{+0.7}_{-1.2}\right)~\textmd{ps}^{-1}$ \cite{DiLuzio:2019jyq}.

%=============================================================================

\subsection{Sensitivity to BSM contributions}
\FloatBarrier
\label{sec:sens-obs}

\begin{table}[t]
    \centering
    \resizebox{\textwidth}{!}{
    \begin{tabular}{cc|c|c}\hline
        Process & Observable & Two-fermion operators & Four-fermion operators \\\hline
        $pp\rightarrow t\bar t$ &   $\sigma^\textmd{inc}$   &   $\tilde C_{uG}$ &  - \\
        $pp\rightarrow t\bar t\gamma$ & $\sigma^\textmd{fid}$ & $\tilde C_{uB}$, $\tilde C_{uW}$, $\tilde C_{uG}$ & - \\
        $pp\rightarrow t\bar tZ$ & $\sigma^\textmd{inc}$ & $\tilde C_{uB}$, $\tilde C_{uW}$, $\tilde C_{uG}$, 
        $\tilde C_{\varphi q}^{-}$, $\tilde C_{\varphi u}$& - \\
        $t\rightarrow b W$ & $F_{0,L}$   &   $\tilde C_{uW}$ & -\\
        Top decay & $\Gamma_t$   &  $\tilde C^{(3)}_{\varphi q}$, $\tilde C_{uW}$ & -\\\hline
        $Z\rightarrow b \bar b$ &   $A_{FB}^b$, $R_b$, $\sigma_\textmd{had}$   &    $\tilde C^+_{\varphi q}$    &   -   \\\hline
        $b\rightarrow s \gamma$ &   BR  & $\left[\tilde C_{uB}\right]$, $\left[\tilde C_{uW}\right]$, $\left\{\tilde C_{uG}\right\}$, $\left[\tilde C_{\varphi q}^{(3)}\right]$ & - \\
        $b\rightarrow s \ell^+\ell^-$ &  \makecell{ BR, $A_\textmd{FB}$, $P^{(\prime)}_i$, \\$S_i$, $F_L$, $d\textmd{BR}/dq^2$}  & $\left[\tilde C_{uB}\right]$, $\left[\tilde C_{uW}\right]$, $\left\{\tilde C_{uG}\right\}$, $\tilde C_{\varphi q}^{+(*)}$, $\left[\tilde C_{\varphi q}^{(3)}\right]$ & $\tilde  C_{lq}^{+(*)}$, $\tilde  C_{qe}^{(*)}$ 
        \\
        $b\rightarrow s \nu\bar\nu$ & BR & 
        $\tilde C_{\varphi q}^{+(**)}$
        &
        $\tilde  C_{lq}^{-(*)}$ 
        \\
        Mixing  &   $\Delta M_s$    &   $\left[\tilde C_{uW}\right]$, $\left\{\tilde C_{uG}\right\}$, $\left[\tilde C_{\varphi q}^{(1)}\right]$, $\left[\tilde C_{\varphi q}^{(3)}\right]$   &  - \\
        \hline
        $e^+e^-\rightarrow t\bar t$ & $\sigma$, $A_\textmd{FB}$ & $\tilde C_{uB}$,
        $\tilde C_{uW}$, $\left\{\tilde C_{uG}\right\}$, 
        $\tilde C_{\varphi q}^{-}$,
        $\tilde C_{\varphi u}$ & $\tilde C_{eu}$, $\tilde C_{qe}$, $\tilde C_{lu}$, 
        $\tilde C_{lq}^{-}$
        \\\hline
    \end{tabular}
    }
    \caption{SMEFT contributions to  the observables included in the fit. Coefficients  without  parentheses arise at tree level. Coefficients marked as $[\tilde C_i]$ contribute only at one-loop level to $B$ physics observables while contributions marked as $\{\tilde C_i\}$ are induced by SMEFT and WET running at $\mathcal{O}(\alpha_s)$ only. Coefficients $\tilde C_i^{(*)}$ and $\tilde C_i^{(**)}$ receive contributions at one-loop level that change their tree-level definitions, see 
    Eqs.\eqref{eq:ast}, \eqref{eq:2ast}.  }
    \label{tab:sensitivity}
\end{table}

In Tab.~\ref{tab:sensitivity} we summarize which linear combinations of SMEFT Wilson coefficients contribute to each observable. 
Contributions denoted in square brackets $[\tilde C_i]$ are induced at one-loop level only, while those written as $\{\tilde C_i\}$ contribute only via RGE evolution.
Tree-level coefficients marked with an asterisk receive additional contribution at one-loop level, which are suppressed by at least one order of magnitude, see 
    Eqs.\eqref{eq:ast},\eqref{eq:2ast} and Appendix~\ref{app:numerics} for details.

Total cross sections of the top-quark production channels, the top-quark decay width, and the helicity fractions measured at the LHC allow to test six coefficients of the operators in Eq.~\eqref{eq:top_boson}, namely, $\tilde C_{uB}$, $\tilde C_{uW}$, $\tilde C_{uG}$, $\tilde C_{\varphi u}$, $\tilde C_{\varphi q}^{(1)}$, and $\tilde C^{(3)}_{\varphi q}$~\footnote{ At the LHC, single top production is sensitive  to these coefficients as well.
However, bin-to-bin correlations are not publicly available and we therefore do not consider these observables, see also footnote 1.}. 
While $t\bar tZ$ production is only sensitive to the linear combination $\tilde C^{-}_{\varphi q}$ (see Eq.~\eqref{eq:def-Cpm}), the total decay width is sensitive to $\tilde C^{(3)}_{\varphi q}$. 
Thus, including data from top-quark decay allows to test $\tilde C^{(1)}_{\varphi q}$ and $\tilde C^{(3)}_{\varphi q}$ individually.
Note that contributions from $\tilde C_{uG}$ to any  of the  $B$-physics and lepton collider observables we consider  arise only from RGE evolution at $\mathcal{O}(\alpha_s)$ and mixing.

Observables of $Z\rightarrow b\bar b$ decay are sensitive to $\tilde C^{+}_{\varphi q}$, and the other operators considered here do not contribute to this process.

Including  $b\rightarrow s$ observables allows to put new and stronger constraints on  SMEFT coefficients. 
The interplay  of $b\rightarrow s\gamma$ transitions with $t \bar t \gamma$ has been worked out in \cite{Bissmann:2019gfc}.
BSM contributions to the former are induced at one-loop level by $\tilde C_{uB}$, $\tilde C_{uW}$, $\tilde C_{uG}$, and $\tilde C^{(3)}_{\varphi q}$.

For $b\rightarrow s \ell^+ \ell^-$ transitions, tree level contributions to $\Delta C_{9,10}$ arise from $\tilde C^+_{\varphi q}$, $\tilde C^+_{lq}$, defined in Eq.~\eqref{eq:def-Cpm}, and $\tilde C_{qe}$. 
The latter cancels, however, in the left-chiral combination $\Delta C_{9}-\Delta C_{10}$, which is the one that
gives the dominant interference term in semileptonic $B$ decays with the SM. We therefore expect only little sensitivity to  $\tilde C_{qe}$ from these modes. On the other hand, this highlights the importance of $B_s \to \mu \mu$, which is sensitive to $C_{10}$ only. 
%}
At one-loop level, all eleven SMEFT operators considered here contribute to $\Delta C_{9,10}$ ($\tilde C_{uG}$ only via mixing). In the case of $\tilde C^{(1)}_{\varphi q}$, $\tilde C_{\varphi u}$, $\tilde C^{(1)}_{lq}$, $\tilde C^{(3)}_{lq}$, $\tilde C_{lu}$, $\tilde C_{qe}$, $\tilde C_{e u}$, and partially $\tilde C^{(3)}_{\varphi q}$, these contributions can simply be absorbed by redefining the fit degrees of freedom
\begin{align}
    \begin{aligned} \label{eq:ast}
    \tilde C^{+(*)}_{lq} &= \tilde{C}^{+}_{lq} + \frac{\alpha }{\pi \sin^2\theta_w}\left(I_1(x_t)\tilde C_{lu} + I_2(x_t)\tilde C^{+}_{lq}\right) 
    \,,\\
    \tilde C^{(*)}_{q e} &= \tilde{C}_{q e} + \frac{\alpha }{\pi \sin^2\theta_w}\left( I_1(x_t)\tilde C_{eu} + I_2(x_t)\tilde C_{q e} \right)\,,\\
    \tilde C^{+(*)}_{\varphi q} &= \tilde C^+_{\varphi q} + \frac{\alpha }{\pi \sin^2\theta_w} \left( I_1(x_t)\tilde C_{\varphi u} + I_2(x_t) \tilde C^{+}_{\varphi q} \right) 
    \,.
    \end{aligned}
\end{align}
Numerically, these loop-level corrections are typically below percent-level  compared to tree-level contributions.
For the remaining contributions from $\tilde C^{(3)}_{\varphi q}$, $\tilde C_{uB}$, $\tilde C_{uW}$ (and $\tilde C_{uG}$) to $\Delta C_{9,10}$ such a redefinition  is not possible and additional degrees of freedom arise. However, these remaining contributions  to $\Delta C_{9,10}$ are at least one order of magnitude smaller than the tree-level ones.

At tree level, $b\rightarrow s \nu\bar \nu$ transitions are sensitive to  $\tilde C^+_{\varphi q} + \tilde C^-_{lq}$.
Additional loop-level contributions by $\tilde C_{uW}$, $\tilde C^{(1)}_{\varphi q}$, $\tilde C^{(3)}_{\varphi q}$, $\tilde C_{\varphi u}$, $\tilde C^{(1)}_{lq}$, $\tilde C^{(3)}_{lq}$ and $\tilde C_{lu}$ can be absorbed into   $\tilde C^{+(**)}_{\varphi q}$ and $\tilde C^{-(*)}_{lq}$:
\begin{align}
    \begin{aligned} \label{eq:2ast}
        \tilde C^{+(**)}_{\varphi q} &= 
        \tilde C^{+}_{\varphi q}-\sqrt{2}\frac{\alpha m_t}{\pi m_W}I_{uW}^{\nu}\tilde C_{uW} -\frac{\alpha}{\pi}I^{\nu(3)}_{\varphi q}(x_t)\tilde C_{\varphi q}^{(3)}+\frac{\alpha}{\pi}\left(I^\nu_{lu}(x_t)\tilde C_{\varphi u}+I^{\nu(1)}_{\varphi q}\tilde C_{\varphi q}^{(1)}\right)\,,\\
      \tilde C^{-(*)}_{lq} &= \tilde C^{-}_{lq}+\frac{\alpha}{\pi}\left(I^\nu_{lu}(x_t)\tilde C_{lu}+I^{\nu(1)}_{lq}(x_t)\tilde C^{(1)}_{lq}\right)+\frac{\alpha}{\pi}I^{\nu(3)}_{lq}(x_t)\tilde C^{(3)}_{lq}\,.
    \end{aligned}
\end{align}
Dineutrino observables depend only on the sum of these  coefficients.
Meson mixing is sensitive at one-loop level to $\tilde C_{uW}$, $\tilde C^{(1)}_{\varphi q}$, and $\tilde C^{(3)}_{\varphi q}$ while contributions from $\tilde C_{uG}$ arise only through SMEFT $\mathcal{O}(\alpha_s)$ RGE evolution.
Electroweak RGE effects in $B$ physics \cite{Feruglio:2017rjo} as well as in top-quark physics are included in the numeric fits but are not shown here for clarity. 

In summary, while all SMEFT coefficients contribute to the $B$ physics observables considered, these effects are mostly induced at one-loop level and thus naturally suppressed. Notable exceptions are tree-level contributions from $\tilde C^{+}_{\varphi q}$, $\tilde C^{+}_{lq}$, $\tilde C_{qe}$, and $\tilde C^{-}_{l q} + \tilde C^{+}_{\varphi q}$. 
In addition, $\tilde C_{uB}$ is important as it contributes  with sizable coefficient  to $\Delta C_7$ \cite{Bissmann:2019gfc}.
Thus, we expect that $B$ physics data  constrains these SMEFT-coefficients rather strongly, and  the others much less.

\begin{figure}[t]
    \centering
    \includegraphics[width=0.6\textwidth]{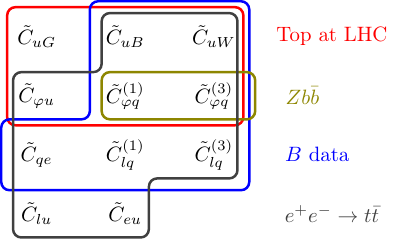}
    \caption{Schematic overview of dominant contributions from  SMEFT coefficients to the different sets of  observables considered in this work.
    Subleading contributions at one-loop level as well as mixing-induced ones from $\tilde C_{uG}$ are neglected.}
    \label{fig:scheme-dof}
\end{figure}

Measurements of top-quark pair production cross sections and the forward-backward asymmetry at a lepton collider are sensitive to four linear combinations of two-fermion operators 
$\tilde C_{uB}$, $\tilde C_{uW}$, $\tilde C^-_{\varphi q}$, and $\tilde C_{\varphi u}$. The sensitivity to $\tilde C_{uG}$ is smaller because contributions 
arise only through RGE evolution.
While these coefficients affect the $ttZ$ and $tt\gamma$ vertex, four-fermion operators can also contribute in following linear combinations: $\tilde C^-_{lq}$, $\tilde C_{qe}$, $\tilde C_{eu}$, and $\tilde C_{lu}$. 
Thus, combining $\ell^+\ell^-\rightarrow t\bar t$ observables with top-quark ones at LHC and $B$ physics observables allows to test the complete 11-dimensional parameter space. In particular, coefficients $\tilde C_{eu}$ and $\tilde C_{lu}$ remain only poorly constrained by Belle II and the HL-LHC.
A summary   of  the dominant contributions to the different observables is illustrated  in Fig.~\ref{fig:scheme-dof}.

%=============================================================================

\section{Fits to present data}
\label{sec:fit-now}

We employ EFT\textit{fitter} \cite{Castro:2016jjv}, which is based on the \textit{Bayesian Analysis Toolkit - BAT.jl} \cite{Schulz:2020ebm}, to constrain the Wilson coefficients in a Bayesian interpretation. 
We include systematic and statistical experimental and SM theory uncertainties. All uncertainties on the measured observables are assumed to be Gaussian distributed.  
The procedure of our fit is detailed in our previous analyses in Refs.~\cite{Bissmann:2019gfc,Bissmann:2019qcd}, and is based on Ref.~\cite{Castro:2016jjv}.

BSM contributions are parametrized as in \eqref{eq:interpol}, which  includes quadratic dimension-six terms. 
While these purely BSM contributions are formally of higher order in the EFT expansion,  $\mathcal{O}(\Lambda^{-4})$, it has been shown \cite{Bissmann:2019qcd, Hartland:2019bjb} that the inclusion of such quadratic terms has only a negligible effect on constraints of coefficients for which the linear term in the EFT expansion gives the dominant contribution, as expected in regions where the EFT is valid.

We include measurements of observables from both top-quark pair production processes and top-quark decay at the LHC, $Z\to b\bar b$ transitions, and $b\rightarrow s$ transitions from different experiments.
Measurements of the same observable from different experiments can in principle be correlated \cite{Aaboud:2019pkc}. 
Correlations are included as long as they are provided, comprising mainly bin-to-bin correlations and correlations between the $W$ boson helicity fractions. 
Unknown correlations can affect the result of the fit significantly \cite{Bissmann:2019qcd}. 
Therefore, we follow a strategy similar to the ones of  Refs.~\cite{Durieux:2019rbz, Brivio:2019ius} and include only the most precise measurement of an observable in the fit. 
Especially, if no complete correlation matrices for differential distributions are given by the experiments, we do not include these measurements in the analysis.
For $B$ physics observables, a variety of measurements have been combined by the Heavy Flavour Averaging Group (HFLAV) \cite{Amhis:2019ckw}. 
Wherever possible, we include their averaged experimental values in our analysis.
For all remaining unknown correlations between different observables, we make the simplifying assumptions that the measurements included in the fit are uncorrelated. 

%=============================================================================

We work out current constraints from top-quark measurements in Sec.~\ref{sec:top-today}, from $Z\to b\bar b$ data in Sec.~\ref{sec:Zbb-con}, from $B$-physics in Sec.~\ref{sec:b-today},
and perform a global analysis in Sec.~\ref{sec:today}.

\subsection{Current constraints from top-quark measurements at the LHC \label{sec:top-today}}

\begin{table}[t]
    \centering
    \begin{tabular}{ccccccc}\hline
         Process    &   Observable  &   $\sqrt{s}$  &   Int. luminosity &  Experiment & Ref. &  SM Ref.  \\\hline
         $t\bar t \gamma$   &   $\sigma^\mathrm{fid}(t\bar{t}\gamma, 1\ell)\,,\ \sigma^\mathrm{fid}(t\bar{t}\gamma, 2\ell)$ &   13~TeV  &   36.1~$\textmd{fb}^{-1}$  &  ATLAS  & \cite{ Aaboud:2018hip} &   \cite{Aaboud:2018hip, Melnikov:2011ta}\\\hline
         $t\bar t Z$ &   $\sigma^\mathrm{inc}(t\bar{t} Z)$ &   13~TeV  &   77.5~$\textmd{fb}^{-1}$   &   CMS  & \cite{CMS:2019too} &  \cite{Frixione:2015zaa, deFlorian:2016spz, Frederix:2018nkq} \\  \hline
         $t\bar t$   &   $\sigma^\textmd{inc}(t\bar t)$ &   13~TeV  &   36.1~$\textmd{fb}^{-1}$    &  ATLAS & \cite{Aad:2019hzw}    &  \cite{Czakon:2011xx}\\
         &   $F_0\,,\ F_L$   &   8~TeV  &   20.2~$\textmd{fb}^{-1}$  &   ATLAS & \cite{Aaboud:2016hsq} & \cite{Czarnecki:2010gb}\\
         & $\Gamma_t$   &   8~TeV   &   20.2~$\textmd{fb}^{-1}$ &   ATLAS   &  \cite{Aaboud:2017uqq}    &   \cite{Gao:2012ja} \\\hline
    \end{tabular}
    \caption{Considered observables for top-quark processes at the LHC and references for the corresponding measurements and SM calculations.}
    \label{tab:top_data}
\end{table}

In Tab.~\ref{tab:top_data} we summarize the measurements and the corresponding SM predictions of the top-quark observables at the LHC included in the fit. 
This dataset comprises measurements of fiducal cross sections $\sigma^\textmd{fid}(t\bar t\gamma,1\ell)$ ($\sigma^\textmd{fid}(t\bar t\gamma,2\ell)$) of $t\bar t\gamma$ production in the single lepton (dilepton) channel, inclusive cross sections $\sigma^\textmd{inc}(t\bar t)$ and $\sigma^\textmd{inc}(t\bar t Z)$ of $t\bar t$ and $t\bar t Z$ production, respectively, measurements of the $W$ boson helicity fractions $F_{0,L}$, and a measurement of the total top-quark decay width $\Gamma_t$. 
The SM predictions for $t\bar t\gamma$ cross sections include NLO QCD corrections Refs.~\cite{Aaboud:2018hip, Melnikov:2011ta}, while predictions for $t\bar tZ$ cross sections are computed at NLO QCD including electroweak corrections \cite{Frixione:2015zaa, deFlorian:2016spz, Frederix:2018nkq}.
For $t\bar t$ production, the SM prediction at NNLO QCD is taken from Ref.~\cite{Aad:2019hzw}, and has been computed following Ref.~\cite{Czakon:2011xx}.
Predictions for helicity fractions and the total decay width include NNLO QCD corrections \cite{Czarnecki:2010gb,Gao:2012ja}. 

\begin{figure}[t]
    \centering
    \includegraphics[width=0.5\textwidth]{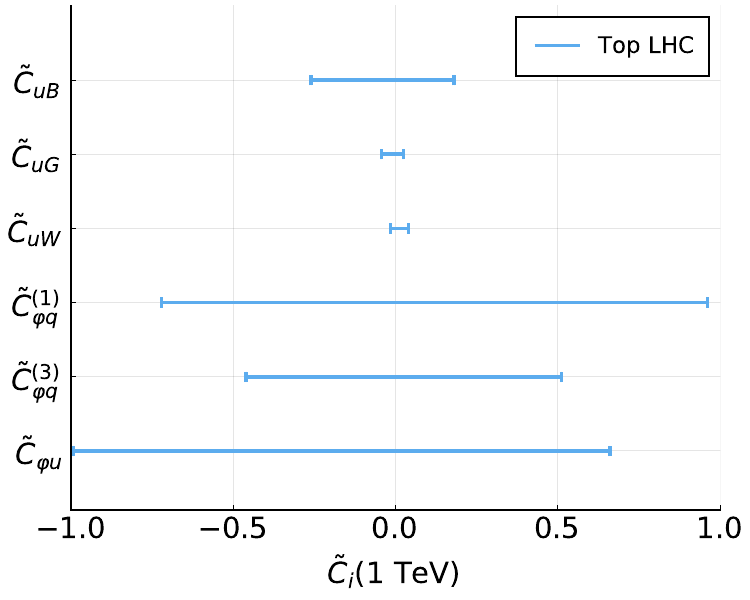}
    \includegraphics[width=0.49\textwidth]{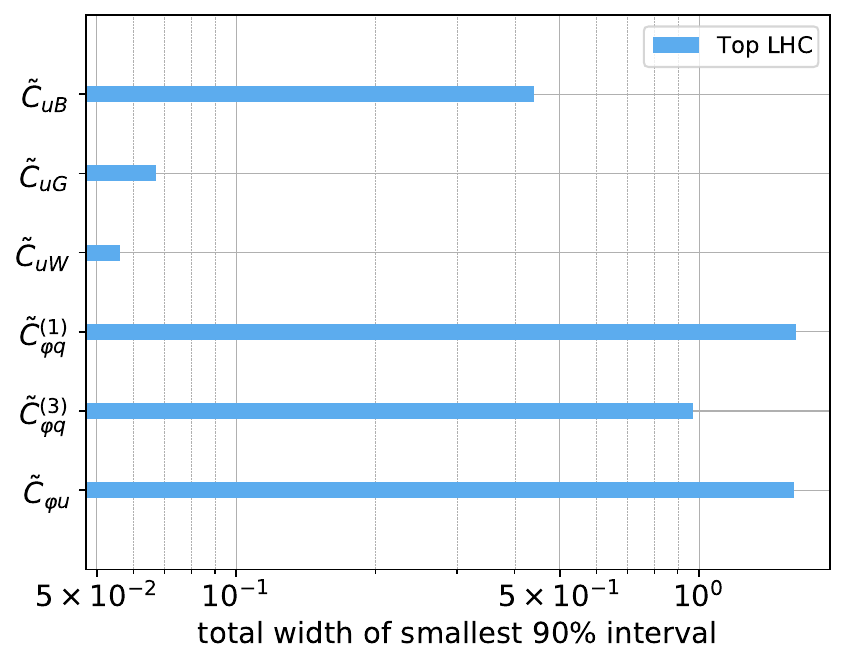}
    \caption{Constraints on SMEFT Wilson coefficients $\tilde C_i$ from the measurements of top-quark observables in Tab.~\ref{tab:top_data}. Shown are the marginalized smallest intervals containing \SI{90}{\percent} posterior probability (left) and the total width of these intervals (right) obtained in a fit of six coefficients to top-quark data.
    For all coefficients we choose a uniform distribution in the interval $-1\leq \tilde C_i \leq 1$ as the prior probability.}
    \label{fig:top-now-1D}
\end{figure}

In Fig.~\ref{fig:top-now-1D} we give constraints on SMEFT Wilson coefficients detailed in Tab.~\ref{tab:sensitivity} obtained in a fit of six coefficients to the data in Tab.~\ref{tab:top_data}.
The strongest constraints are found for $\tilde C_{uG}$ and $\tilde C_{uW}$, which are at the level of $\mathcal{O}(10^{-2})$ and stem from measurements of $t\bar t$ production cross sections and the $W$ boson helicity fractions, respectively.
Constraints on $\tilde C_{uB}$, which are dominated by $t\bar t\gamma$ measurements, are at the level of $\mathcal{O}(10^{-1})$. 
Including measurements of the top-quark decay width allows constraining $\tilde C^{(3)}_{\varphi q}$ to a level of $\mathcal{O}(1)$. 
Given present data, results are limited by large experimental uncertainties.
Both $\tilde C^{(1)}_{\varphi q}$ and $\tilde C_{\varphi u}$ remain almost unconstrained by the measurements of $t\bar t Z$ production due to a strong correlation between their contributions and larger uncertainties of measurements and theory predictions.

%=============================================================================
\subsection{Constraints from \boldmath$Zbb$ measurements}
\label{sec:Zbb-con}

Precision measurements of $Z$ pole observables have been performed at LEP 1 and SLC, and the results are collected in Ref.~\cite{Zyla:2020zbs}.
In our analysis, we focus on those that are sensitive to BSM contributions which affect the $Zb\bar b$ vertex.
The measurements included are those of the forward-backward asymmetry and the ratio of partial widths for $Z\rightarrow f\bar f$  \cite{ALEPH:2005ab} 
\begin{align}
    A^{0,b}_{FB}{}^\textmd{Exp} = 0.0996\pm0.0016\,,\quad     R_b{}^\textmd{Exp}=0.21629\pm0.00066\,.
\end{align}
The corresponding SM values are given as \cite{ALEPH:2005ab,Zyla:2020zbs}
\begin{align}
    A^{0,b}_{FB}{}^\textmd{SM} = 0.1030\pm0.0002\,,\quad R_b^\textmd{SM} = 0.21581\pm0.00002\,.
\end{align}
These observables are sensitive to BSM contributions from $\tilde C^{+}_{\varphi q}$, which alter the $Zb\bar b$ vertex, and allow to derive strong constraints on this coefficient. 
\begin{figure}[t]
    \centering
    \includegraphics[width=0.7\textwidth]{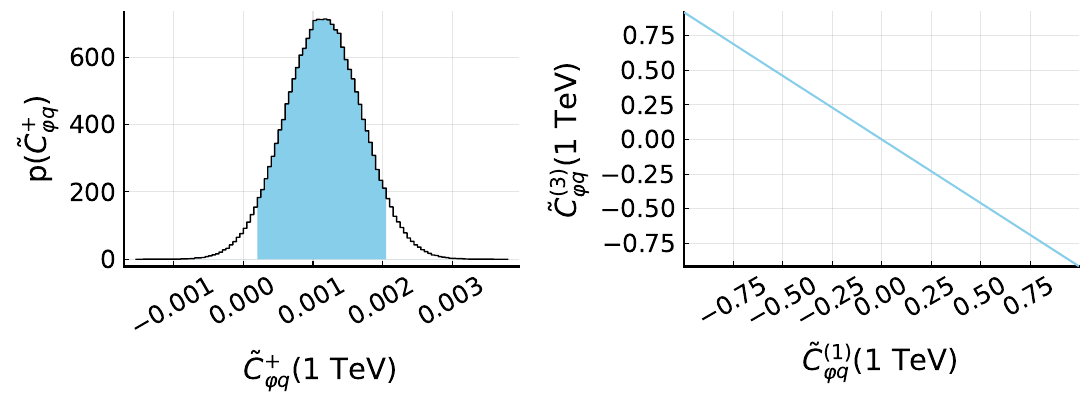}
    \caption{Results of fits to $Zb\bar b$ data considering $\tilde C^{+}_{\varphi q}$ (left) and $\tilde{C}^{(1)}_{\varphi q}$ and $\tilde{C}^{(3)}_{\varphi q}$ (right) as degrees of freedom. Shown are the one-dimensional (left) and two-dimensional (right) projection of the posterior distribution obtained in fits of one and two coefficients, respectively. Colored areas correspond to the smallest intervals containing 90\,\% of the posterior distribution.
    For the prior we consider an interval $-1\leq \tilde C_i\leq 1$.
    }
    \label{fig:Zbb-only}
\end{figure}
The results of a fit of one ($\tilde C^+_{\varphi q}$) and two ($\tilde C^{(1)}_{\varphi q}$, $\tilde C^{(3)}_{\varphi q}$) coefficients to $Zb\bar b$ data are shown in Fig.~\ref{fig:Zbb-only}.
As can be seen, this dataset strongly constrains $\tilde C^{+}_{\varphi q}$ to a level of $\mathcal{O}(10^{-3})$. 
Due to the deviations from the SM present in $A^{0,b}_{FB}$ we observe deviations of about $2\,\sigma$ in $\tilde C^{+}_{\varphi q}$.
Considering results in the $\tilde{C}^{(1)}_{\varphi q}$-$\tilde{C}^{(3)}_{\varphi q}$ plane we find, as expected, strong correlations, and only a very small slice of the two-dimensional parameter space is allowed by present data.

\subsection{Current constraints from \boldmath$B$ physics measurements  \label{sec:b-today}}

 \begin{table}[h]
    \centering
    \begin{tabular}{cccccc}\hline
        Process    &   Observable  & $q^2$ bin [GeV$^2$] & Experiment &   Ref. &  SM Ref.  \\\hline
        $\bar{B}\rightarrow X_s \gamma$   &   BR$_{E_\gamma>1.6~\textmd{GeV}}$ & - &   HFLAV    & \cite{Amhis:2019ckw} &   \cite{Misiak:2015xwa}\\
        $B^0\rightarrow K^* \gamma$   &   BR & - &   HFLAV    & \cite{Amhis:2019ckw} &   \cite{Straub:2018kue}\\
        $B^+\rightarrow K^{*+} \gamma$   &   BR & - &   HFLAV    & \cite{Amhis:2019ckw} &   \cite{Straub:2018kue}\\
         \hline
        $\bar{B}\rightarrow X_s \ell^+\ell^-$  &   BR & ${[1,6]}$ &   BaBar  & \cite{ Lees:2013nxa} & \cite{Huber:2015sra} \\
        &   $A_\textmd{FB}$ & ${[1,6]}$ &   Belle  & \cite{Sato:2014pjr}  & \cite{Huber:2015sra} \\
        $B_s \rightarrow \mu^+\mu^-$    &   BR  &   -   &   LHCb&   \cite{LHCb2021}   &   \cite{Straub:2018kue}    \\
        ${B}^0\rightarrow K^* \mu^+\mu^-$  &   \makecell{$F_L\,,\ P_1\,,\ P_2\,,\ P_3\,,$\\$\ P_4^\prime\,,\ P_5^\prime\,,\ P_6^\prime\,,\ P_8^\prime $} & ${[1.1,6]}$ &   LHCb & \cite{ Aaij:2020nrf} & \cite{Straub:2018kue} \\
        ${B}^0\rightarrow K \mu^+\mu^-$  &   $d \text{BR}/d q^2$ & ${[1,6]}$ &   LHCb & \cite{ Aaij:2014pli} & \cite{Straub:2018kue} \\
        ${B}^+\rightarrow K^+ \mu^+\mu^-$  &   $d \text{BR}/d q^2$ & ${[1,6]}$ &   LHCb & \cite{ Aaij:2014pli} & \cite{Straub:2018kue} \\
        ${B}^+\rightarrow K^{+*} \mu^+\mu^-$  &   $d \text{BR}/d q^2$ & ${[1,6]}$ &   LHCb & \cite{ Aaij:2014pli} & \cite{Straub:2018kue} \\
        ${B}_s\rightarrow \phi \mu^+\mu^-$  &   $F_L\,,\ S_3\,,\ S_4\,,\ S_7$ & ${[1,6]}$ &   LHCb & \cite{ Aaij:2015esa} & \cite{Straub:2018kue} \\
        ${\Lambda}_b\rightarrow \Lambda \mu^+\mu^-$  &   $d \text{BR}/d q^2$ & ${[15,20]}$ &   LHCb & \cite{ Aaij:2015xza} & \cite{Straub:2018kue} \\
         \hline
        $B_s-\bar B_s$ mixing   &   $\Delta M_s$    &   -   &   HFLAV   &   \cite{Amhis:2019ckw}    &   \cite{DiLuzio:2019jyq}\\\hline
    \end{tabular}
    \caption{$B$-physics measurements included in the fit. For observables measured in $q^2$ bins (where $q^2$ denotes the squared invariant dilepton mass) we include only one bin due to unknown correlations between different bins for consistency.  
    }
    \label{tab:B_data}
\end{table}

In Tab.~\ref{tab:B_data} we give the $B$ physics observables and the corresponding references of the measurements and SM predictions considered in our fit.
This dataset includes both inclusive and exclusive branching ratios of $b\rightarrow s\gamma $ transitions, total and differential branching ratios of various $b\rightarrow s\mu^+\mu^-$ processes, inclusive branching ratios and asymmetries of $b\rightarrow s\ell^+\ell^-$ transitions, and angular distributions of $B^0\to K^* \mu^+\mu^-$ and $B_s \to \phi \mu^+\mu^-$. 
In the case of $B_s\to \mu^+\mu^-$, we consider the latest results presented by the LHCb collaboration \cite{LHCb2021}.
We compute SM predictions and uncertainties with \texttt{flavio} \cite{Straub:2018kue}.
In addition, we also include the mass difference $\Delta M_s$ measured in $B_s-\bar B_s$ mixing, with  SM prediction from Ref.~\cite{DiLuzio:2019jyq}.
Note that we do not take into account measurements of the $B\rightarrow K^{(*)}\nu\bar\nu$ branching ratios as only upper limits are presently available by Belle~\cite{Lutz:2013ftz} and BaBar~\cite{Lees:2013kla}, which can not be considered in EFT\textit{fitter}.

\begin{figure}[t]
    \centering
    \includegraphics[width=0.5\textwidth]{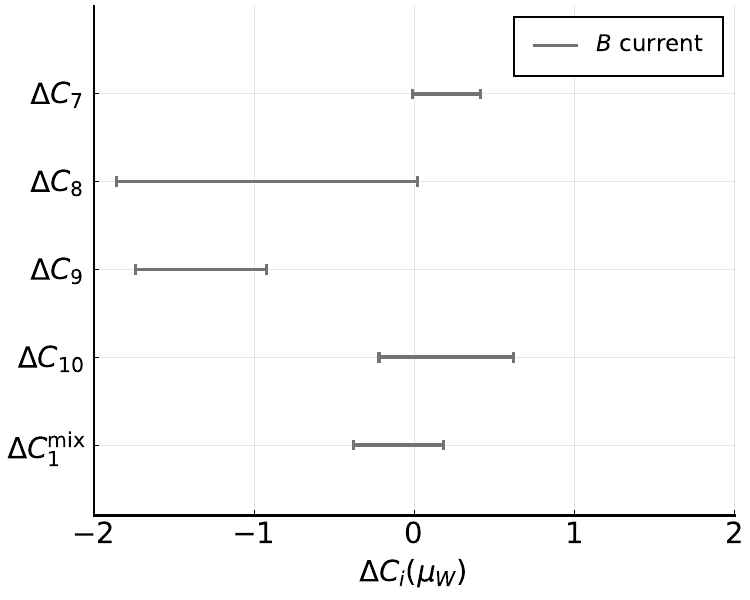}
    \includegraphics[width=0.49\textwidth]{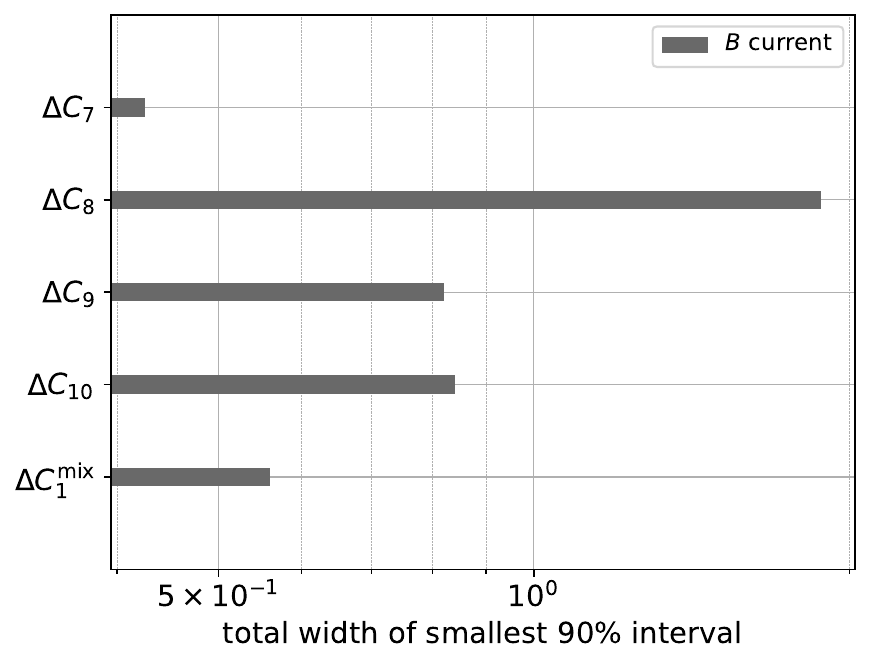}\\
    \includegraphics[width=0.5\textwidth]{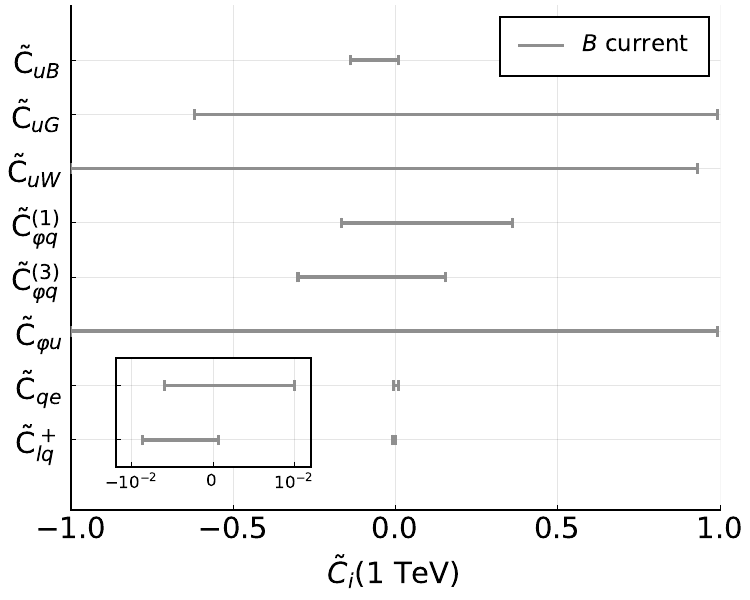}
    \includegraphics[width=0.49\textwidth]{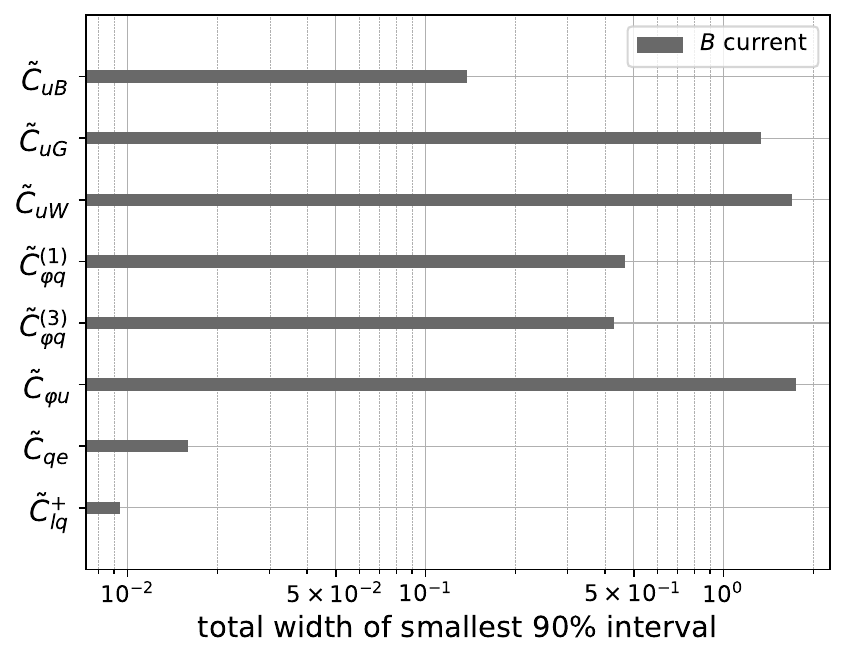}
    \caption{Constraints on WET coefficients $\Delta C_i$ at the scale $\mu=\mu_W$ (upper plots) and on SMEFT coefficients (lower plots) from measurements of $B$ observables in Tab.~\ref{tab:B_data}. Shown are the marginalized smallest intervals containing \SI{90}{\percent} of the posterior probability (left) and the total width of these intervals (right) obtained in a fit of five WET (upper plots) and eight SMEFT coefficients (lower plots) to the data.
    The fit is performed using a uniform distribution over the interval $-2\leq \Delta C_i \leq 2$ ($-1\leq \tilde C_i \leq 1$) as a prior for the WET (SMEFT) coefficients.}
    \label{fig:B-now-WET-1D}
\end{figure}

In Fig.~\ref{fig:B-now-WET-1D} (upper plots) we give constraints on BSM contributions $\Delta C_i$ to WET coefficients at the scale $\mu_W=m_W$ from a fit of five coefficients to the data in Tab.~\ref{tab:B_data}. 
The strongest constraints exist for $\Delta C^\textmd{mix}_1$ and $\Delta C_7$ for which the width of the smallest \SI{90}{\percent} interval is about $5.5\times 10^{-1}$ and $ 4\times 10^{-1}$, respectively.
The weakest constraints are obtained for $\Delta C_8$ as this coefficient contributes via mixing only.
For $\Delta C_9$ we observe deviations from the SM. This effect stems mainly from measurements of angular distributions of $b\rightarrow s\mu^+\mu^-$ by LHCb and is widely known and discussed in literature, see e.g. Ref.~\cite{Aebischer:2018iyb} for a detailed discussion.
The exact deviation from the SM depends on the measurements considered in the fit. For the observables in Tab.~\ref{tab:B_data} we find deviations mostly in $\Delta C_9$ while $\Delta C_{10}$ is SM like. 
The constraints on the WET coefficients $\Delta C_i$ can be translated into constraints on eight SMEFT coefficients (Fig.~\ref{fig:B-now-WET-1D}, lower plots) discussed in more detail in the next subsection, which are strongly correlated due to the matching conditions, see Eqs.~\eqref{eq:match-tree}-\eqref{Eq:MatchC1}.
Nevertheless, strong constraints at the level of $\mathcal{O}(10^{-2})$ are found for the four-fermion coefficients. Constraints on the remaining coefficients are around one ($\tilde C_{uB}$, $\tilde C^{(1)}_{\varphi q}$, $\tilde C^{(3)}_{\varphi q}$) to two ($\tilde C_{uG}$, $\tilde C_{uW}$, $\tilde C_{\varphi u}$) orders of magnitude weaker.
Note that deviations from the SM, which are present in the one-dimensional projection of the posterior distribution of $\Delta C_9$, can not be seen in the one-dimensional results in the SMEFT basis. 
This is due to the strong correlations among the SMEFT coefficients induced by the matching conditions.

\subsection{Combined fit to current data \label{sec:today}}

\begin{figure}[t]
    \centering
    \includegraphics[width=0.49\textwidth]{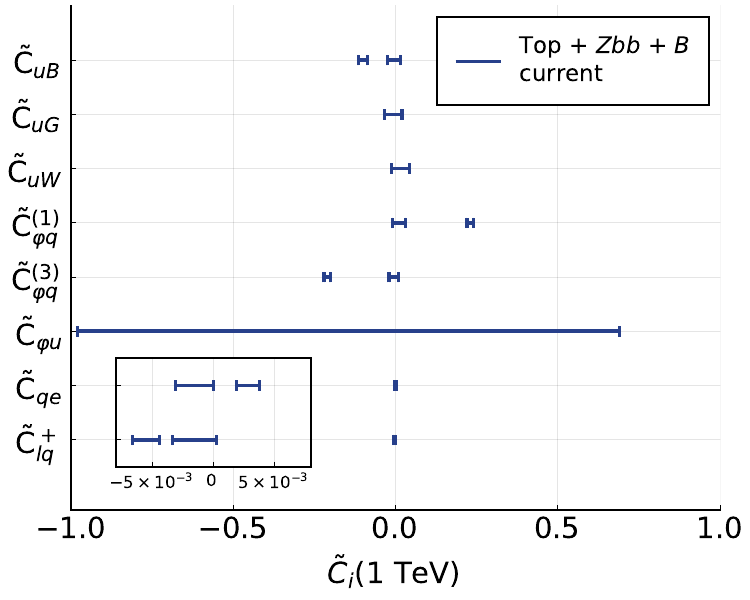}
    \includegraphics[width=0.49\textwidth]{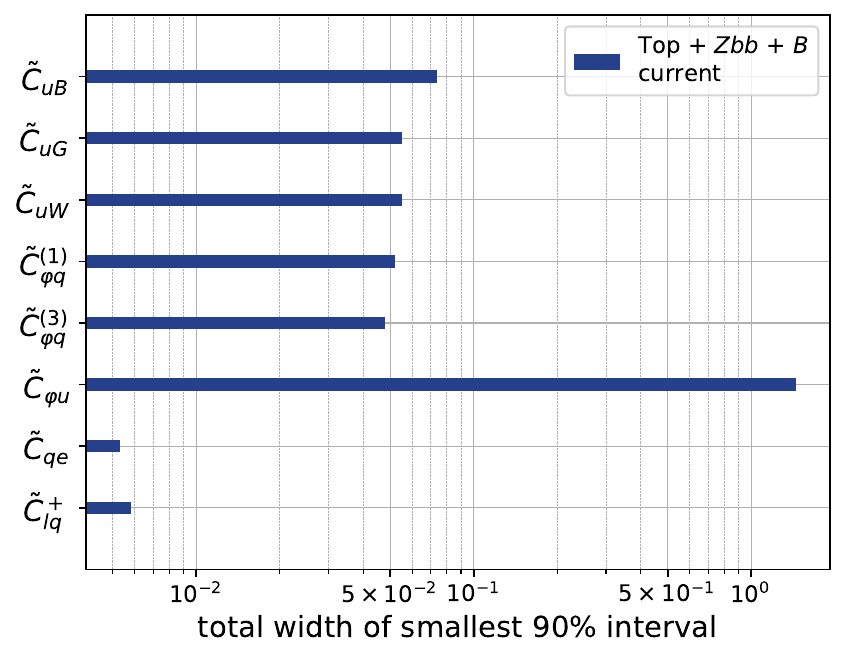}
    \caption{Constraints on SMEFT coefficients $\tilde C_i$ in Eq.~\eqref{eq:max-dof} from a fit of eight coefficients to top-quark data in Tab.~\ref{tab:top_data}, $Zbb$ data, and $B$ physics data in Tab.~\ref{tab:B_data}. Shown are the smallest intervals containing \SI{90}{\percent} posterior probability (left) and the total width of these intervals (right). 
    For the prior we assume a uniform distribution over the interval $-1\leq \tilde C_i\leq 1$.
    } 
    \label{fig:TB-Now-maxDof}
\end{figure}

Combining top-quark, $Zbb$, and $B$ observables allows to constrain a larger number of SMEFT coefficients compared to fits using only the individual datasets.
Specifically, the  coefficients constrained by data in Tabs.~\ref{tab:top_data} and \ref{tab:B_data} and $Zbb$ data are
\begin{align}
    \tilde C_{uB}\,,\quad \tilde C_{uG}\,,\quad \tilde C_{uW}\,,\quad \tilde C_{\varphi q}^{(1)}\,,\quad \tilde C_{\varphi q}^{(3)}\,,\quad  \tilde C_{\varphi u}\,,\quad \tilde C_{qe}\,,\quad \tilde C_{lq}^+\,.
    \label{eq:max-dof}
\end{align}
From the fit of these eight coefficients to the combined dataset we obtain the results shown in Fig.~\ref{fig:TB-Now-maxDof}.
\begin{figure}[t]
    \centering
    \includegraphics[width=0.49\textwidth]{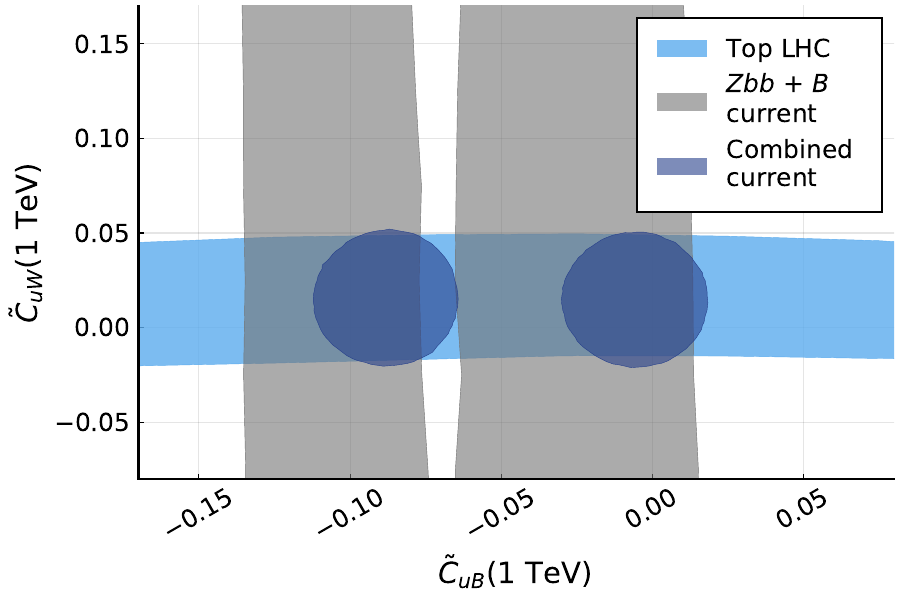}
    \includegraphics[width=0.49\textwidth]{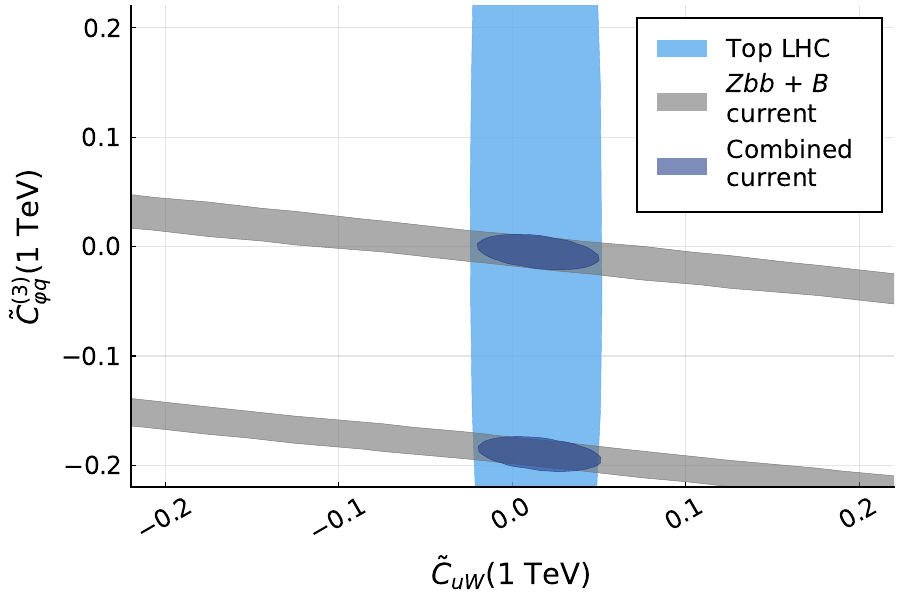}
    \caption{Examples for two-dimensional posterior distributions of SMEFT coefficients $\tilde C_i$ in Eq.~\eqref{eq:max-dof} obtained in a fit of eight coefficients to top-quark data (light blue), $B$ physics data (grey) and the combined dataset including $Zbb$ data (blue). 
    Shown are the smallest intervals containing \SI{90}{\percent} of the posterior distribution.  
    For the prior we assume a uniform distribution over the interval $-1\leq \tilde C_i\leq 1$.
    } 
    \label{fig:TB-overlay-examples}
\end{figure}
The strongest constraints are on $\tilde C_{qe}$ and $\tilde{C}^+_{lq}$, whose width of the smallest interval is around $(5-6)\times10^{-3}$.
This is expected, since both coefficients give sizable contributions to $\Delta C_9$ and $\Delta C_{10}$ at tree level \eqref{eq:match-tree}.
For $\tilde C_{uB}$, $\tilde C_{uG}$, $\tilde{C}_{uW}$, $\tilde C^{(1)}_{\varphi q}$ and $\tilde C^{(3)}_{\varphi q}$ the constraints are about one order of magnitude weaker, with a width of around $(5-7)\times 10^{-2}$.
While constraints on $\tilde C_{uG}$ and $\tilde{C}_{uW}$ coincide with those derived from fits to top-quark data, the combination of the three datasets significantly tightens constraints on the other three coefficients.
For $\tilde{C}_{uB}$ this enhancement stems from different sensitivities of top-quark and $B$ data, as already discovered in Ref.~\cite{Bissmann:2019gfc}. 
The effect of the different datasets is shown in detail in Fig.~\ref{fig:TB-overlay-examples} (left), where we give the two-dimensional projection of the posterior distributions obtained in fits to different datasets in the $\tilde C_{uB}$-$\tilde C_{uW}$ plane. 
Here, the effects are even more pronounced compared to Ref.~\cite{Bissmann:2019gfc}, since a larger set of $B$ observables is considered here.
Constraints on $\tilde C^{(1)}_{\varphi q}$ and $\tilde C^{(3)}_{\varphi q}$ (Fig.~\ref{fig:TB-Now-maxDof}) are tightened by the inclusion of $Zbb$ data, which strongly constraints $\tilde C^+_{\varphi q}$, as well as the strong constraints on $\tilde C^{(3)}_{\varphi q}$, which arise from the combination of top-quark and $B$ physics data (see Fig.~\ref{fig:TB-overlay-examples}).
As can be seen, in the combined fit the SM is included in the smallest intervals containing 90\,\% of the posterior distribution of $\tilde C^{(1)}_{\varphi q}$ and $\tilde C^{(3)}_{\varphi q}$, which is shown in detail in Fig.~\ref{fig:combined-now-1D-Deviations}.
The weakest constraints are found for $\tilde C_{\varphi u}$, since contributions to $B$ physics data are strongly suppressed, and $t\bar tZ$ production offers only a limited sensitivity, as we can already see in Fig.~\ref{fig:top-now-1D}.

Interestingly, we find two solutions for several coefficients; one of which is SM like, while the other one deviates from the SM: $\tilde C_{uB}$, $\tilde C^{(1)}_{\varphi q}$, $\tilde C^{(3)}_{\varphi q}$, and the four-fermion coefficients $\tilde C_{qe}$ and $\tilde C^+_{lq}$. 
As can be seen in Fig.~\ref{fig:TB-overlay-examples}, the second solutions stem from the correlations between the coefficients introduced by matching the SMEFT basis onto the WET basis. 
Since the number of degrees of freedom is smaller in WET, correlations among the coefficients arise.
Inclusion of top-quark data reduces these correlations, however, for the five coefficients the sensitivity of top-quark observables does not suffice to exclude the non-SM branches completely given present data and theory predictions.
%In general, the quadratic ansatz for BSM contributions \eqref{eq:interpol}, which in principle allows for two solutions. 
Without further input this ambiguity cannot be resolved.

We compare our results to those reported in a recent study on  $b\rightarrow s\ell^+\ell^-$ transitions \cite{Ciuchini:2020gvn}.
In contrast to our analysis, operators are defined in a basis of diagonal down-type quark Yukawa couplings, which leads to an additional factor of
$1/(V_{tb}V_{ts}^*)$. Taking this factor into account, the results from  \cite{Ciuchini:2020gvn} correspond to 
$\tilde C^{+}_{lq}, \tilde C_{qe} \sim 10^{-3}$, consistent with Fig.~\ref{fig:TB-Now-maxDof}.  
Repeating our fit with $\tilde C_{qe}$ and $\tilde C^+_{lq}$ only, we find agreement with Ref.~\cite{Ciuchini:2020gvn}.

We also comment on Drell-Yan production at the LHC. 
Amongst the couplings with  top-quark focus considered in this works, (\ref{eq:future-SMEFT-coff}), this concerns
$\tilde C_{\varphi q}^+$, $\tilde C_{qe}$ and $\tilde C_{lq}^+$, just like $b\to s \ell^+ \ell^-$ and $Z \to b \bar b$.
Drell-Yan limits 
from existing data  and a $3000 \, \mbox{fb}^{-1}$ future projection 
for the  semileptonic four-fermion operators with $b$-quarks
are at the  level of $\mathcal{O}(10^{-2})$ 
\cite{Greljo:2017vvb, Fuentes-Martin:2020lea}, and weaker than in the combined fit,  Fig.~\ref{fig:TB-Now-maxDof}.
Note,  with the flavor of the initial quarks in $pp$-collisions undetermined an actual measurement of a quark flavor-specific coefficient is  not possible.
A detailed study of the implications of Drell-Yan processes for a global fit is beyond the scope of this work.

%=============================================================================

\section{Impact of future colliders}
\label{sec:fit-future}

Both the HL-LHC operating at $14~$TeV with an  integrated luminosity of $3000~$fb$^{-1}$ \cite{Atlas:2019qfx} and Belle II at 
$50~$ab$^{-1}$ \cite{Kou:2018nap} are going to test the SM at the next level of precision.
In Sec.~\ref{Sec:Exp-Precision}, we work out the impact of  future measurements at these facilities on the 
SMEFT Wilson coefficients.

A first study of top-quark physics at the proposed lepton collider CLIC has been provided in Ref.~\cite{Abramowicz:2018rjq}. 
CLIC is intended to operate at three different center-of-mass energies: 380~GeV, 1.4~TeV, and 3~TeV and two different beam polarizations are foreseen by the accelerator design: a longitudinal polarization of $\pm 80\,\%$ for the electron beam and no polarization of the positron beam. 
We investigate the impact of measurements with the currently foreseen precision of such a lepton collider on the constraints of SMEFT Wilson coefficients
in Sec.~\ref{sec:clic}.

We combine existing data with HL-LHC, Belle II and CLIC  projections in Sec.~\ref{sec:future-combined}.

\subsection{Expected constraints from HL-LHC and Belle II}
\label{Sec:Exp-Precision}

For the expected experimental uncertainties at the HL-LHC and Belle II we adopt estimates of the expected precision by ATLAS, CMS and Belle II collaborations \cite{ATLAS:2018yyd,Atlas:2019qfx,CMS:2018clr,CMS:2018vga,Kou:2018nap}. 
If no value for the systematic uncertainties is given, we assume that these uncertainties shrink by a factor of two compared to the current best measurement, which is the case for the $t\bar t$ and $t\bar t Z$ cross sections, the $W$ boson helicity fractions, and the top-quark decay width.
In addition, we make the assumption that theory uncertainties shrink by a factor of two compared to the current SM uncertainties due to improved MC predictions and higher-order calculations. 
We summarize the observables and references for the expected experimental and theory precision at HL-LHC and Belle II in Tab.~\ref{tab:HLLHC-BelleII}.
For the purpose of the fit, we consider present central values of measurements for the future projections. 
If no measurement is available, we consider the SM for central values.

For fiducial cross sections of $t \bar t \gamma$ production, an analysis with the expected uncertainties is provided in Refs.~\cite{ATLAS:2018yyd,Atlas:2019qfx}. For both the dilepton and single-lepton cross section we consider the precision of the channel with the largest experimental uncertainty as our estimate. 
For $t\bar tZ$ production we follow the analysis in Refs.~\cite{CMS:2018clr,Atlas:2019qfx} and scale statistical uncertainties according to the luminosity. For systematic uncertainties we assume for simplicity a reduction by a factor of 2. For estimating the expected precision of the total $t \bar t$ production cross section, we base our assumptions on the study of differential $t\bar t$ cross sections in Ref.~\cite{CMS:2018vga,Atlas:2019qfx}. 
For the uncertainties we apply the same assumptions as for $t\bar tZ$. 
As the $W$ boson helicity fractions and the top-quark decay width are not discussed in Ref.~\cite{Atlas:2019qfx}, we treat them in the same way as the $t\bar t$ cross section for simplicity.

\begin{table}[t]
    \centering
    \resizebox{\textwidth}{!}{
    \begin{tabular}{cccccc}\hline
        Process    &   Observable  & $q^2$ bin [GeV$^2$] &  Experiment & Ref. &  SM Ref.  \\\hline
        $t\bar t \gamma$   &   $\sigma^\mathrm{fid}(t\bar{t}\gamma, 1\ell)\,,\ \sigma^\mathrm{fid}(t\bar{t}\gamma, 2\ell)$ & - &  ATLAS  & \cite{ATLAS:2018yyd,Atlas:2019qfx} &   \cite{Aaboud:2018hip, Melnikov:2011ta}\\
        $t\bar t Z$ &   $\sigma^\mathrm{inc}(t\bar{t} Z)$ & - &   CMS  & \cite{CMS:2018clr,Atlas:2019qfx} &  \cite{Frixione:2015zaa, deFlorian:2016spz, Frederix:2018nkq} \\  \hline
        $t\bar t$   &   $\sigma^\textmd{inc}(t\bar t)$  & - & CMS & \cite{CMS:2018vga,Atlas:2019qfx}    &  \cite{Czakon:2011xx}\\
        &   $F_0\,,\ F_L$   & - &  - & - & \cite{Czarnecki:2010gb}\\
        &   $\Gamma_t$   & - &  - & - & \cite{Gao:2012ja}\\\hline
        $\bar{B}\rightarrow X_s \gamma$   &   BR$_{E_\gamma>1.6~\textmd{GeV}}$ & - &   Belle II    & \cite{Kou:2018nap} &   \cite{Misiak:2015xwa}\\
        $B^0\rightarrow K^* \gamma$   &   BR & - &   Belle II    & \cite{Kou:2018nap} &   \cite{Straub:2018kue}\\
        $B^+\rightarrow K^{+*} \gamma$   &   BR & - &   Belle II    & \cite{Kou:2018nap} &   \cite{Straub:2018kue}\\
         \hline
        $\bar{B}\rightarrow X_s \ell^+\ell^-$  &   BR, $A_\textmd{FB}$  & ${[3.5,6]}$ &   Belle II  & \cite{Kou:2018nap} & \cite{Huber:2015sra} \\
        ${B}^0\rightarrow K^* \mu^+\mu^-$  &   \makecell{$F_L$\,, $P_1\,,\ P_2\,,\ P_3\,,$\\$P_4^\prime\,,\ P_5^\prime\,,\ P_6^\prime\,,\ P_8^\prime $} & \makecell{${[1.1,2.5]}$, ${[2.5,4]}$, ${[4,6]}$} &   Belle II & \cite{ Kou:2018nap} & \cite{Straub:2018kue} \\   \hline  
        ${B}^0\rightarrow K^{(*)} \nu\bar\nu$  &   BR  &    -   &  Belle II & \cite{ Kou:2018nap} & \cite{Straub:2018kue} \\\hline
    \end{tabular}
        }
    \caption{Overview of observables considered at future HL-LHC and Belle II projections. For each process we denote the references for the experimental projection and the SM prediction. 
    In case of the [1.1,2.5] $q^2$ bin for ${B}^0\rightarrow K^* \mu^+\mu^-$ we consider the Belle II projection of the [1.0,2.5] bin for the expected experimental uncertainties. 
    }
    \label{tab:HLLHC-BelleII}
\end{table}

For measurements of $b\rightarrow s$ transitions we take the estimates in Ref.~\cite{Kou:2018nap} into account. For the $b\rightarrow s\gamma$ inclusive branching ratio we take the precision for the BR($\bar{B}\rightarrow X_s\gamma)_{E_\gamma>1.9 ~\textmd{GeV}}$ measurement and assume that the same uncertainties 
apply for $E_\gamma>1.6 ~\textmd{GeV}$. In case of $B^{(+)}\rightarrow K^{(+)*}\gamma$, we directly include the estimated precision in Ref.~\cite{Kou:2018nap}. Similarly, for the inclusive decay $\bar B\rightarrow X_s \ell^+\ell^-$ we use the expected precision for the $3.5~\textmd{GeV}^2\leq q^2\leq 6~\textmd{GeV}^2$ bin. We also considered other bins for this observable and found very comparable sensitivity.
Finally, for $B\rightarrow K^*\mu^+\mu^-$ we include the angular distribution observable $P^{(\prime)}_i$ in different $q^2$ bins, and study the implications of the anomalies found in present data of $b\rightarrow s\mu^+\mu^-$ angular distributions.

\begin{figure}[t]
    \centering
    \includegraphics[width=0.5\textwidth]{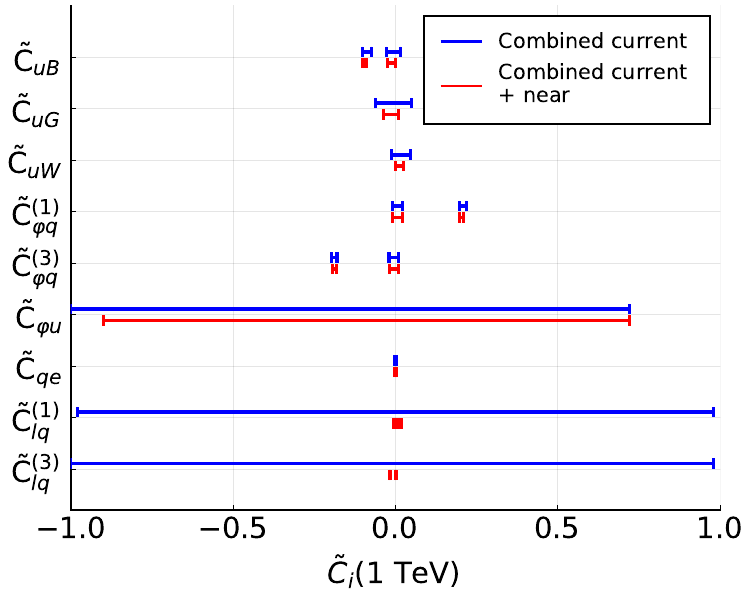}
    \includegraphics[width=0.49\textwidth]{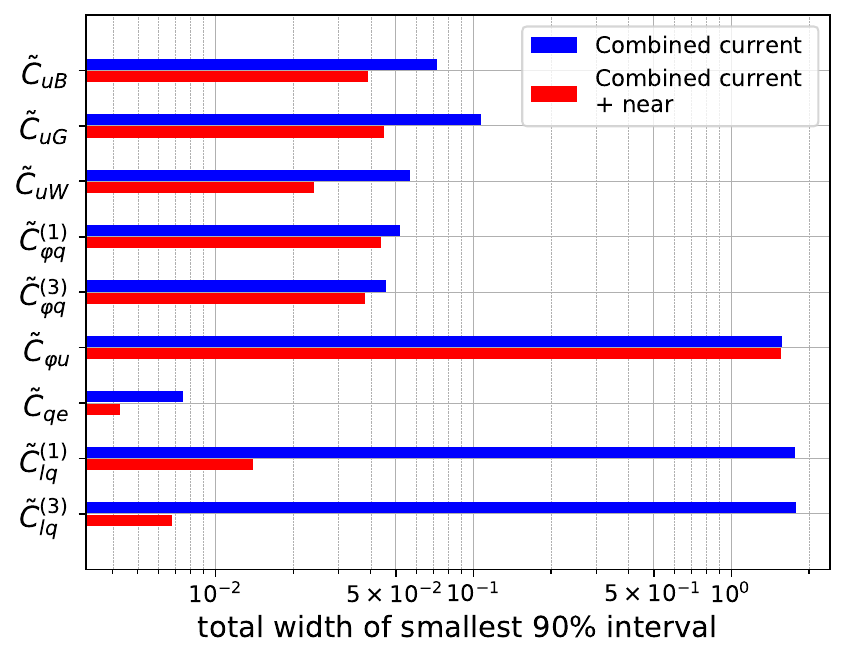}
    \caption{Constraints on coefficients $\tilde C_i$ from fits of nine coefficients to current top-quark and $B$ measurements in Tabs.~\ref{tab:top_data} and \ref{tab:B_data} (blue) and to current measurments and projections of top-quark and $B$ observables in Tabs.~\ref{tab:top_data}-\ref{tab:HLLHC-BelleII} (red). 
    Shown are the marginalized smallest intervals containing \SI{90}{\percent} posterior probability (left) and the total widths of these intervals (right).}
    \label{fig:TopB-near-1D}
\end{figure}
\begin{figure}[t]
    \centering
    \includegraphics[width=0.5\textwidth]{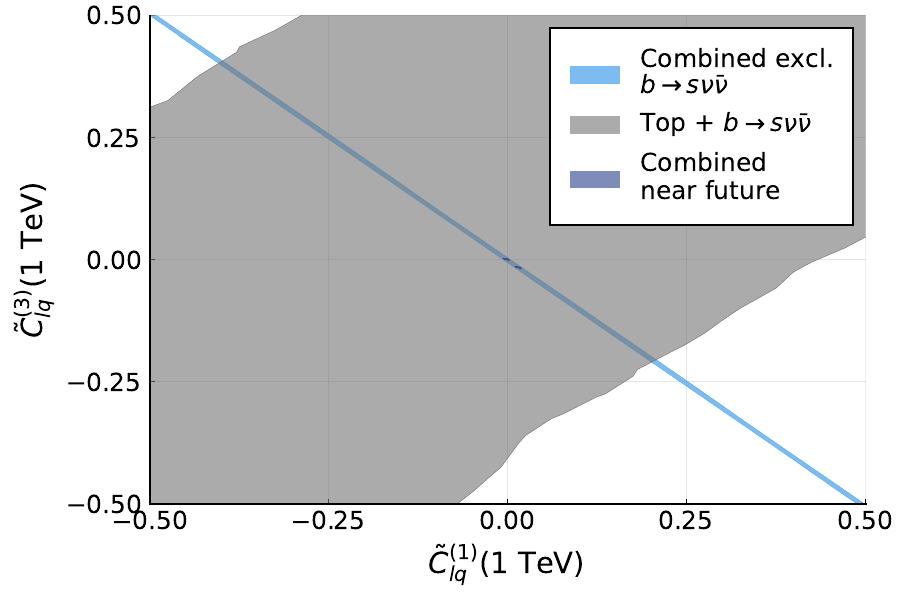}
    \caption{Two-dimensional projection of the posterior distribution in the $\tilde C^{(1)}_{lq}$-$\tilde C^{(3)}_{lq}$ plane.
    Shown are smallest intervals containing \SI{90}{\percent} of the posterior distribution obtained in fits of nine coefficients in the near future projection combining top-quark observables with all beauty observables except $b\rightarrow s\nu\bar\nu$ ones (light blue), only with $b\rightarrow s\nu\bar\nu$ ones (grey), and the combined set (dark blue).
    }
    \label{fig:TopB-near-overlay}
\end{figure}
Combining top-quark and $B$ observables at HL-LHC and Belle II allows to test a total of nine SMEFT coefficients, see Fig.~\ref{fig:TopB-near-1D}. 
In order to derive these constraints with EFT\textit{fitter}, we have chosen a smaller prior $|\tilde C_i|\leq 0.1$ for the four-fermion coefficients because the posterior distribution lies only in a very small region, and a larger prior would lead to convergence issues.
At this point, we neglect subleading contributions from $\tilde C_{eu}$ and $\tilde C_{lu}$, which are considered in Sec.~\ref{sec:future-combined}.
As can be seen, the observables strongly constrain all coefficients except for $\tilde C_{\varphi u}$, which is only very weakly constrained, $\mathcal{O}(1)$, due to the low sensitivity in both $t\bar tZ$ and $B$ observables.
Conversely, the strongest constraints are found for the four-fermion coefficients, around $4\times 10^{-3}$ and $(7-15)\times 10^{-3}$ for $\tilde C_{qe}$ and $\tilde C^{(1,3)}_{lq}$, respectively.
The inclusion of $b\rightarrow s\nu\bar\nu$ observables allows to test $\tilde C^{(1)}_{lq}$ and $\tilde C^{(3)}_{lq}$ independently due to the orthogonal sensitivity compared to $b\rightarrow s\ell^+\ell^-$ observables, as indicated in Fig.~\ref{fig:TopB-near-overlay}.
We observe that the interval obtained in the combined fit is significantly smaller than expected from the simple overlay of constraints from $b\rightarrow s\nu\bar\nu$ and $b\rightarrow s\ell^+\ell^-$ observables.
The reason is, that the posterior distribution is constrained in the multi-dimensional hyperspace, and the combination significantly reduces correlations among different coefficients.
In addition, we find that two solutions for $\tilde C^{(1)}_{lq}$ and $\tilde C^{(3)}_{lq}$ are allowed: one is close to the SM, while the other is around $\tilde C^{(1)}_{lq}\sim-\tilde C^{(3)}_{lq}\sim 10^{-3}$, and deviates strongly from the SM.
Without further input, this ambiguity can not be resolved.
Constraints on the remaining coefficients $\tilde C_{uB}$, $\tilde C_{uG}$, $\tilde C_{uW}$, $\tilde C^{(1)}_{\varphi q}$, and $\tilde C^{(3)}_{\varphi q}$ are in the range $(2-5)\times 10^{-2}$. 
Here, the higher precision in the near-future scenario tightens constraints on $\tilde C_{uB}$ ($t\bar t\gamma$ and $b\rightarrow s\gamma$), $\tilde C_{uG}$ ($t\bar t$), and $\tilde C_{uW}$ (helicity fractions) by a factor of around 2, while constraints on $\tilde C^{(1,3)}_{\varphi q}$ remain mostly unchanged.    
Note that the inclusion of additional measurements in the near-future projection does not suffice to resolve the second solutions observed for several coefficients.

\subsection{CLIC projections  \label{sec:clic}}

In Tab.~\ref{tab:top_CLIC} we list the top-quark observables for the CLIC future projections considered in this work.
This set comprises total cross sections of $t\bar t$ production and forward-backward asymmetries $A_\textmd{FB}$ as observables for different energy stages and beam polarizations \cite{Abramowicz:2018rjq}. 
We use the current SM predictions as nominal values, which include NLO QCD corrections \cite{Durieux:2018tev}.

\begin{figure}[t]
    \centering
    \includegraphics[width=0.5\textwidth]{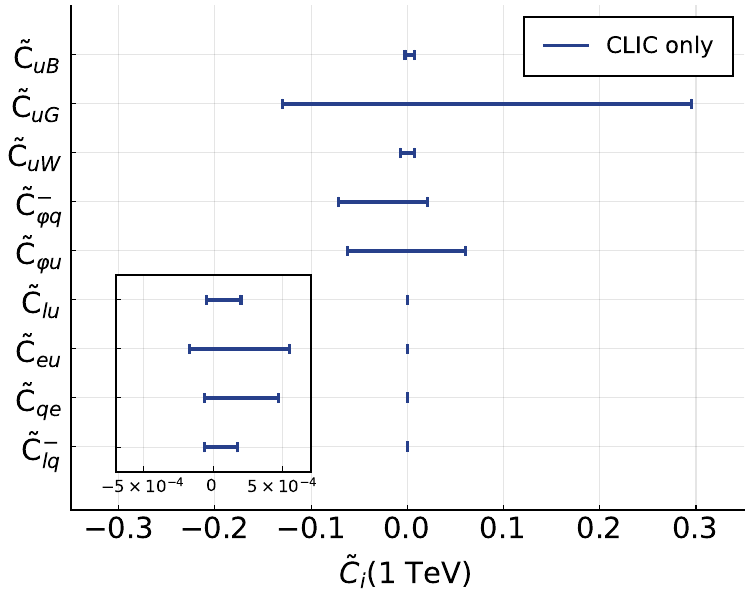}
    \includegraphics[width=0.49\textwidth]{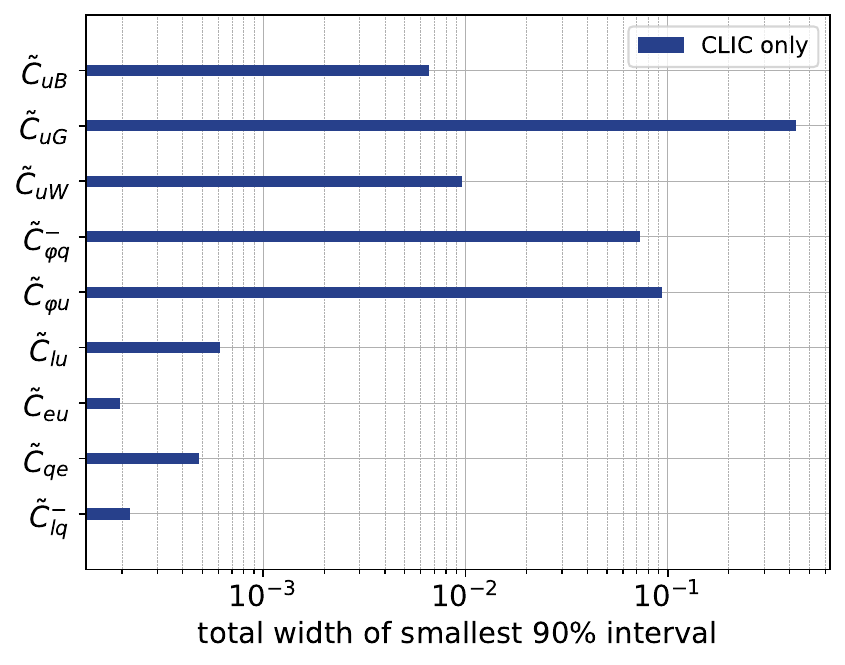}
    \caption{Constraints on coefficients $\tilde C_i$ from a fit of nine coefficients to CLIC observables in Tab.~\ref{tab:top_CLIC}. 
    Shown are the marginalized smallest intervals containing \SI{90}{\percent} posterior probability (left) and the total widths of these intervals (right).}
    \label{fig:Top-far-1D}
\end{figure}

In Fig.~\ref{fig:Top-far-1D} we give the results for a fit to the CLIC projections in Tab.~\ref{tab:top_CLIC}. 
A smaller prior $|\tilde C_i|\leq 0.1$ is employed for the four-fermion coefficients due to the small size of the posterior distribution.
We explicitly checked that we do not remove any solutions.
Constraints on $\tilde C_{uG}$, which contributes via mixing only, are at the level of $4\times 10^{-1}$, and weaker compared to the ones on the remaining Wilson coefficients.
For $\tilde C^{-}_{\varphi q}$ and $\tilde C_{\varphi u}$ the width of the smallest \SI{90}{\percent} interval is at the level of $10^{-1}$. 
In comparison, constraints on $\tilde C_{uB}$ and $\tilde C_{uW}$ are found to be stronger by one order of magnitude.
Even tighter constraints are obtained for four-fermion interactions, where the width of the smallest interval is at the level of $(2-6)\times 10^{-4}$.
\begin{table}[t]
    \centering

    \begin{tabular}{ccccc}\hline
        Observable  &   $\sqrt{s}$ &   Polarization ($e^-,e^+$)   &  Ref. experiment  & SM Ref.   \\\hline
        $\sigma_{t\bar t}$, $A_\textmd{FB}$   & 380~GeV &  $(\pm\SI{80}{\percent},0)$ &   \cite{Abramowicz:2018rjq}  & \cite{Durieux:2018tev}  \\
        $\sigma_{t\bar t}$, $A_\textmd{FB}$   & 1.4~TeV &  $(\pm\SI{80}{\percent},0)$ &   \cite{Abramowicz:2018rjq}  & \cite{Durieux:2018tev}\\
        $\sigma_{t\bar t}$, $A_\textmd{FB}$   &  3~TeV &  $(\pm\SI{80}{\percent},0)$ &   \cite{Abramowicz:2018rjq} &  \cite{Durieux:2018tev} \\\hline
    \end{tabular}
	\caption{Observables at different energies and polarizations for $t\bar t$ production at CLIC \cite{Abramowicz:2018rjq}. 
	SM predictions are taken from \cite{Durieux:2018tev}.
    }
    \label{tab:top_CLIC}
\end{table}
Interestingly, while Fig.~\ref{fig:Top-far-1D} shows results of a fit treating $\tilde C^{-}_{lq}$ as a degree of freedom, the inclusion of RGE effects on $\tilde C^{(1)}_{lq}$ and $\tilde C^{(3)}_{lq}$ allows to distinguish both coefficients. 
The reason is that they develop differently under the RGE flow where corrections are at the level of $\mathcal{O}(1)\,\%$. 
Thus, both coefficients can be constrained simultaneously to a level of $\mathcal{O}(10^{-2})$, as shown in detail in Fig.~\ref{fig:TopB-far-1D}.

\subsection{Combined fit}
\label{sec:future-combined}
Combining measurements and near-future projections of top-quark physics and $B$ physics in Tabs.~\ref{tab:top_data}-\ref{tab:HLLHC-BelleII} with the projections for top-quark observables at a CLIC-like lepton collider allows to constrain all eleven SMEFT coefficients considered in this analysis.

\begin{figure}[t]
    \centering
    \includegraphics[width=0.5\textwidth]{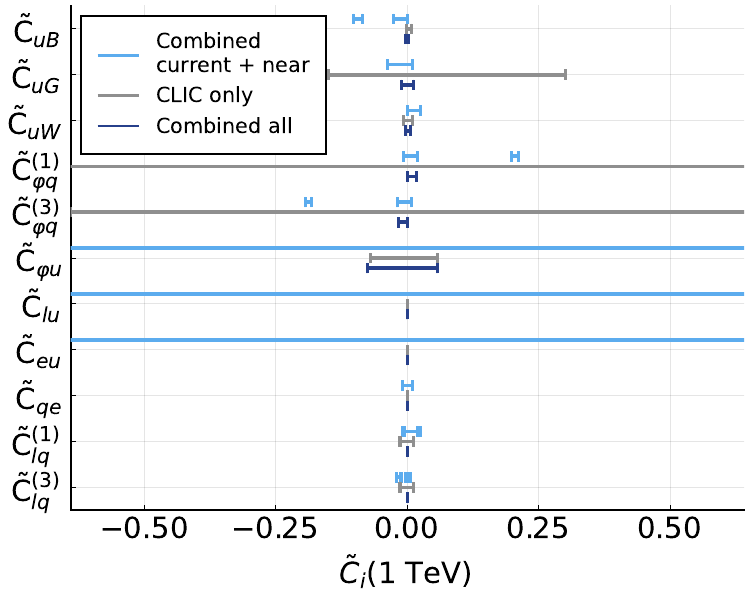}
    \includegraphics[width=0.49\textwidth]{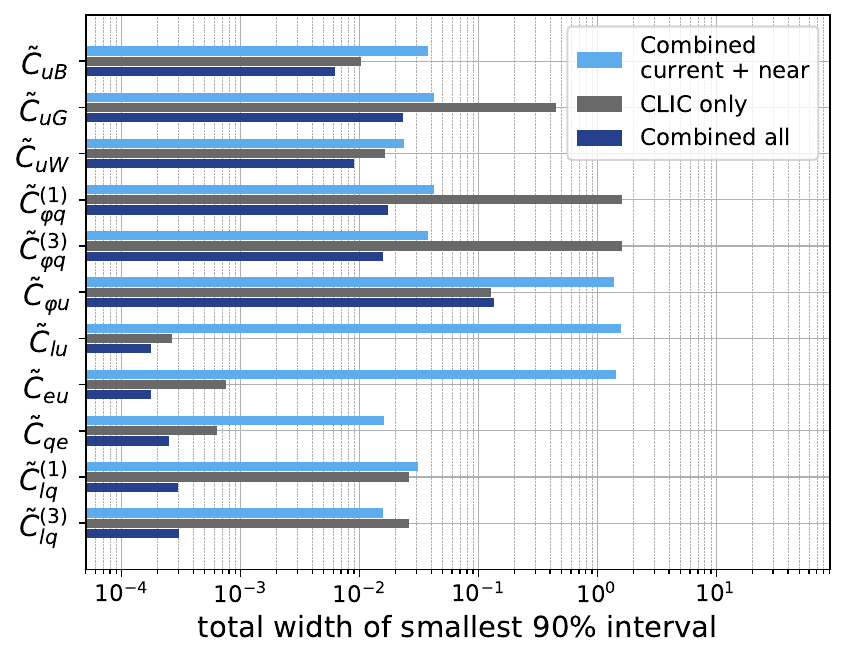}
    \caption{Constraints on coefficients $\tilde C_i$ obtained fitting eleven coefficients to top-quark and $B$ data and near-future projections at HL-LHC and Belle II in Tabs.~\ref{tab:top_data}-\ref{tab:HLLHC-BelleII} and CLIC future projections in Tab.~\ref{tab:top_CLIC}. 
    Shown are the marginalized smallest intervals containing \SI{90}{\percent} posterior probability (left) and the total widths of these intervals (right).
    }
    \label{fig:TopB-far-1D}
\end{figure}

In Fig.~\ref{fig:TopB-far-1D} we give results from fits of all eleven coefficients to current data (Tabs.~\ref{tab:top_data}, \ref{tab:B_data}, and $Zb\bar b$ data) and near-future projections (Tab.~\ref{tab:HLLHC-BelleII}) (light blue), to CLIC projections for top-quark observables (Tab.~\ref{tab:top_CLIC}) (grey) and the combined set (blue).
It can be observed that the fit to the combined set of observables allows to constrain all eleven SMEFT Wilson coefficients.
Flat directions in the parameter space of the coefficients are removed in the global fit.
The strongest constraints are obtained for the four-fermion operators and are at the level of $\mathcal{O}(10^{-4})$.
Constraints on the other operators are weaker and at the level of $\mathcal{O}(10^{-1})$ for $\tilde C_{\varphi u}$ and  $\mathcal{O}(10^{-2})$ for the remaining coefficients. 

\begin{figure}[t]
    \centering
    \includegraphics[width=0.49\textwidth]{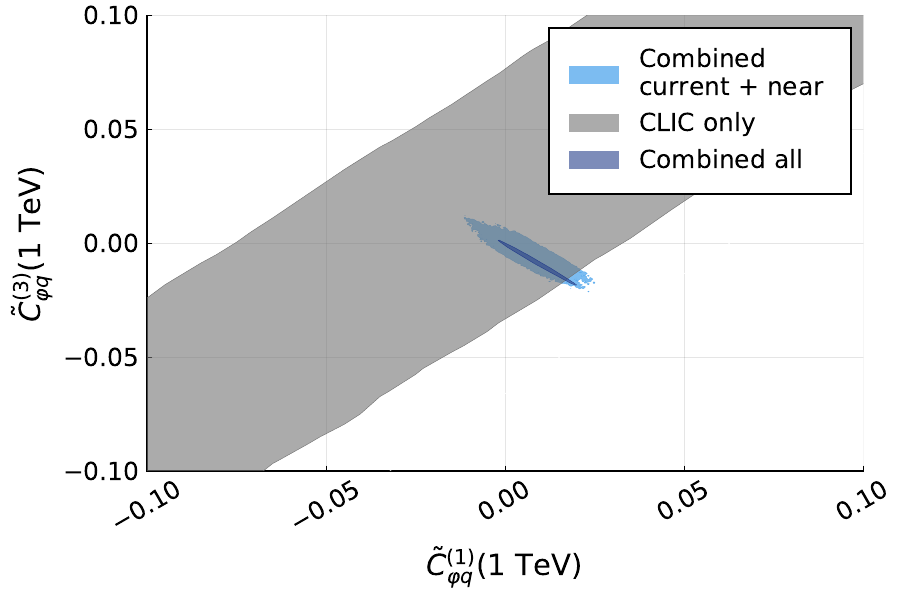}
    \includegraphics[width=0.49\textwidth]{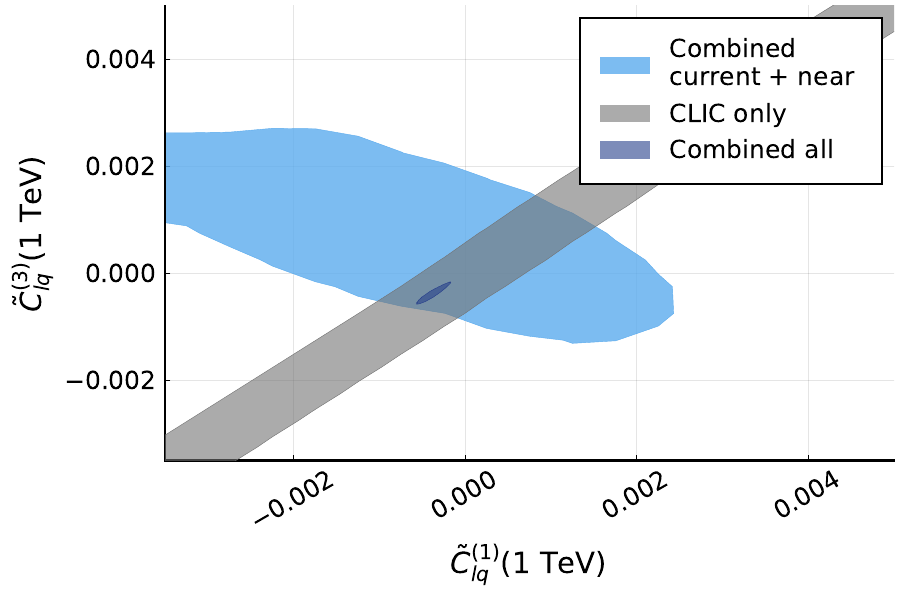}
    \caption{Examples for two-dimensional marginalized posterior distributions of SMEFT coefficients $\tilde C_i$ obtained in a fit of all eleven coefficients to top-quark and $B$ data in Tabs.~\ref{tab:top_data}-\ref{tab:HLLHC-BelleII} (light blue), top-quark observables at CLIC in Tab.~\ref{tab:top_CLIC} 
    (grey) and the combined set (blue). Shown are the smallest regions containing \SI{90}{\percent} posterior probability.
    Constraints from the fit on the combined set are so strong that the smallest \SI{90}{\percent} region is barely visible in the $\tilde C^{(1)}_{lq}$-$C^{(3)}_{lq}$ plane (plot to the right).} 
    \label{fig:Far-overlay-examples}
\end{figure}

As learned previously, combining  different sets of observables yields stronger constraints on all coefficients.
In the case of $\tilde C_{\varphi q}^{(1)}$ and $\tilde C_{\varphi q}^{(3)}$, which are already strongly constrained by present data and Belle II and HL-LHC projections, additional constrains derived from CLIC projections are orthogonal to those from the remaining observables, see Fig.~\ref{fig:Far-overlay-examples} (left).
This tightens the constraints by a factor of two and allows to exclude the second solution. 
Similarly, the second solution for $\tilde C_{uB}$ still present in the near-future scenario is removed as well.
The improvement is particularly significant for $\tilde C^{(1)}_{l q}$ and $\tilde C^{(3)}_{l q}$.
While $b \to s \ell^+ \ell^-$  and $b\rightarrow s \nu\bar \nu$ observables allow to test both coefficients simultaneously, the inclusion of CLIC observables is mandatory to remove the second solution, see Fig.~\ref{fig:Far-overlay-examples} (right).
Correlations, which are induced by CLIC observables, between both coefficients are still present, and sizable deviations from the SM can be found, which is shown in more detail in Fig.~\ref{fig:combined-far-1D-Deviations} in App.~\ref{app:far}. 
These deviations stem from the assumption that Belle II confirms present LHCb data. 
Interestingly, even though CLIC observables strongly constrain $\tilde C^{-}_{lq}$ (assuming that the SM value is measured), the exact position of the smallest 90\,\% interval on the $\tilde C^{(1)}_{\ell q}\sim - \tilde C^{(3)}_{\ell q}$ subspace is determined by Belle II results (Fig.~\ref{fig:combined-far-1D-Deviations}).
A scenario, in which we assume SM values for Belle II observables, is shown in Fig.~\ref{fig:combined-far-1D-noDeviations} in App.~\ref{app:far}, and we find agreement with the SM in this case.
While indeed constraints from CLIC projections and top-quark and $B$ data and projections in the near-future scenario have a different sensitivity, the \SI{90}{\percent} region for $\tilde C^{(1)}_{lq}$ and $\tilde C^{(3)}_{lq}$ is significantly smaller than expected by simply overlaying the constraints obtained in fits to the two individual datasets. 
The reason is that constraints are combined in the full eleven-dimensional hyperspace, and Fig.~\ref{fig:Far-overlay-examples} only shows two-dimensional projections.

As anticipated in Sec.~\ref{sec:SMEFT-SSB} the full, global fit results  including CLIC projections  are obtained assuming lepton-flavor universality. 
While in  BSM scenarios  where  lepton generations couple differently the  results  cannot be applied directly,
the findings on the orthogonality of the constraints and synergies between top and beauty continue to hold.

%=============================================================================

\section{Conclusions}
\label{sec:conclusion}

We performed fits within SMEFT  to  top-quark pair production, decay, $Z\to b\bar b$ transitions, and $b\rightarrow s$ transitions.
We highlight how each of the individual datasets constrains different sets of Wilson coefficients of dimension-six operators affecting top-quark physics at present and future colliders. 
Extending previous works  \cite{Bissmann:2019gfc}, we  put an emphasis on semileptonic four-fermion operators, which are of high interest  as they may be anomalous
according to current flavor  data and moreover are essentially unconstrained for top quarks.
$SU(2)_L$ invariance leads to relations between  up-type and down-type quark observables, a well-known feature with recent applications in semileptonic processes 
within  SMEFT \cite{Bause:2020auq}.
Here, we exploit this symmetry link  between top and beauty observabes at the LHC and a future lepton collider.

Using existing data in Tabs.~\ref{tab:top_data} and \ref{tab:B_data} as well as $Zbb$ data we constrain eight SMEFT Wilson coefficients with results shown in Fig.~\ref{fig:TB-Now-maxDof}.
Combining complementary constraints significantly improves the fits compared to using individual datasets alone, see Fig.~\ref{fig:TB-overlay-examples}.
Going beyond existing data, we entertain a near-future scenario with measurements from Belle II and the HL-LHC,
and one with an additional lepton collider.
While measurements of top-quark observables at the HL-LHC allow to put stronger constraints on the same set of coefficients already tested by present top-quark measurements, 
a notable qualitative improvement  in the near future is the expected observation of $b \to s \nu \bar \nu$ transitions at Belle II, which together with lepton universality
allows to probe four-fermion operators in new ways: orthogonal to $b \to s \ell^+ \ell^-$ and very similar as in contact interactions of leptons and top quarks, see Fig.~\ref{fig:comp}.
Thus, in this near-future scenario a combined fit would allow to probe nine  SMEFT coefficients with estimated precision shown in Fig.~\ref{fig:TopB-near-1D}. 
Combining the present data and projections for near-future experiments together with projections for a CLIC-like lepton collider,  a combined fit enables to constrain the  eleven SMEFT coefficients considered in this work, see Eq.~(\ref{eq:future-SMEFT-coff}), as shown in Fig.~\ref{fig:TopB-far-1D}.
The second solution for $\tilde C^{(1)}_{l q}$ and $\tilde C^{(3)}_{l q}$ present in fits in the near-future scenario, see Fig.~\ref{fig:TopB-near-1D},  could be removed by lepton collider measurements, as demonstrated in  Fig.~\ref{fig:Far-overlay-examples}.
We stress that a lepton collider allows to probe the coefficients $\tilde C_{\varphi u}$, $\tilde C_{l u}$ $\tilde C_{e u}$, which would otherwise remain loosely constrained in the fit.
In the combined fit, constraints on four-fermion coefficients are obtained at the level of $\mathcal{O}(10^{-4})$.

To conclude, in order to extract the strongest constraints on SMEFT coefficients from a global fit of the SMEFT top-quark sector and of relevance to the $b$-anomalies, different collider setups as well as $SU(2)_L$ relations have to be employed to remove flat directions  and to test all possible sources of BSM contributions simultaneously.
The present study clearly demonstrates the strong new physics impact of a future lepton collider.

\smallskip

{\bf Note added:} During the finalization of this project a preprint appeared by CMS  in which
SMEFT coefficients are constrained by top production in association with leptons at the $\sqrt{s}=13$ TeV LHC with $41.5 \, \mbox{fb}^{-1}$ \cite{Sirunyan:2020tqm}.
The constraints on  four-fermion coefficients $\tilde C_{qe}$ and $\tilde C^-_{lq}$ are more than one order of magnitude weaker compared to ours using current data, Fig.~\ref{fig:TB-Now-maxDof}.
However, the CMS-analysis is sensitive to $\tilde C_{eu}$, 
$\tilde C_{lu}$, otherwise unconstrained by present data.
A study of the future physics potential of this  type of analysis would be desirable, however, requires detector-level  simulations and is beyond the scope of this work.

\section*{Acknowledgements}
We thank Danny van Dyk and Susanne Westhoff for useful discussions on RGE
evolution in SMEFT and  Christoph Bobeth for comments on matching SMEFT and WET.
C.G.  is  supported  by  the  doctoral  scholarship program of the \emph{Studienstiftung des deutschen Volkes}. 

\appendix

\section{Weak effective theory}
\label{sec:WET}
At energies below the scale $\mu_W\sim m_W$, physical processes are described by the Weak Effective Theory (WET). 
All BSM particles which are heavier than $m_W$ as well as the top quark and the $W$, $Z$ and Higgs bosons are integrated out. 
Both $b\rightarrow s \ell^+\ell^-$ and $b\rightarrow s \gamma$ transitions are described by the following Lagrangian:
\begin{align}
    \mathcal{L}_\text{WET}^{bs}=\frac{4G_F}{\sqrt{2}}V_{ts}^*V_{tb}\sum_{i=1}^{10} C_i(\mu)Q_i(\mu)\,.
    \label{eq:L_WET}
\end{align}
Here, $G_F$ is the Fermi-constant, $ C_i$ are Wilson coefficients and $Q_i$ are the corresponding effective operators which are defined as follows:
\begin{align}
    \begin{aligned}
        Q_1&=(\bar s_L\gamma_\mu T^ac_L)(\bar c_L\gamma^\mu T^a b_L)\,,
        &Q_2&=(\bar s_L\gamma_\mu c_L)(\bar c_L\gamma^\mu  b_L)\,,\\
        Q_3&=(\bar s_L\gamma_\mu b_L)\sum_q(\bar q\gamma^\mu q)\,,
        &Q_4&=(\bar s_L\gamma_\mu T^ab_L)\sum_q(\bar q\gamma^\mu T^a q)\,,\\
        Q_5&=(\bar s_L\gamma_\mu\gamma_\nu\gamma_\sigma b_L)\sum_q(\bar q\gamma^\mu\gamma^\nu\gamma^\sigma q)\,,
        &Q_6&=(\bar s_L\gamma_\mu\gamma_\nu\gamma_\sigma T^ab_L)\sum_q(\bar q\gamma^\mu\gamma^\nu\gamma^\sigma T^aq)\,,\\
        Q_7&=\frac{e}{16\pi^2}m_b(\bar s_L\sigma^{\mu\nu}b_R)F_{\mu\nu}\,,
        &Q_8&=\frac{g_s}{16\pi^2}m_b(\bar s_L\sigma^{\mu\nu}T^ab_R)G^a_{\mu\nu}\,,\\
        Q_9^{ij}&=\frac{e^2}{16\pi^2}(\bar s_L\gamma_\mu b_L)(\bar\ell^i \gamma^\mu\ell^j)\,,
        &Q_{10}^{ij}&=\frac{e^2}{16\pi^2}(\bar s_L\gamma_\mu b_L)(\bar\ell^i\gamma^\mu\gamma_5\ell^j)\,,
    \end{aligned}
    \label{eq:WET_ops}
\end{align}
with chiral left (right) projectors $L$ ($R$) and the field strength tensor of the photon $F_{\mu\nu}$. We denote charged leptons with $\ell$ and neglect contributions proportional to the subleading CKM-matrix element $V_{ub}$ and to the strange-quark mass.

The effective Lagrangian for  $b \rightarrow s \bar \nu \nu $ transitions can be written as
\begin{align}
     \mathcal{L}_\text{WET}^{ \nu}=\frac{4G_F}{\sqrt{2}}V_{ts}^*V_{tb}\sum_{i,j=1}^3\left( C_L^{ij}(\mu)Q_L^{ij}(\mu) + C_R^{ij}(\mu)Q_R^{ij}(\mu)\right)\,,
    \label{eq:L_nunu}
\end{align}
with effective operators 
\begin{align}
    \begin{aligned}
        Q_L^{ij}=\frac{e^2}{16\pi^2}(\bar s_L\gamma_\mu b_L)(\bar\nu^i \gamma^\mu(1-\gamma_5)\nu^j)\,,\quad
        Q_{R}^{ij}=\frac{e^2}{16\pi^2}(\bar s_R\gamma_\mu b_R)(\bar\nu^i \gamma^\mu(1-\gamma_5)\nu^j)\,.
    \end{aligned}
    \label{eq:nunu_ops}
\end{align}
Assuming flavor universality, only diagonal terms $i=j$ contribute, and all three flavors couple with identical strength.
The $B_s-\bar B_s$ mass difference $\Delta M_s$ can be described as
\begin{align}
     \mathcal{L}_\text{WET}^\textmd{mix}=\frac{G_F^2m_W^2}{16\pi^2}Q_1^\textmd{mix} \left|V_{tb}V_{ts}^*\right|^2 C_{1,tt}^\textmd{mix}\,,
    \label{eq:L_mix}
\end{align}
with the effective operator
\begin{align}
    Q_1^\textmd{mix}= \left(\bar s_L \gamma_\mu  b_L \right)\left(\bar s_L \gamma^\mu  b_L \right)\,.
    \label{eq:mix_ops}
\end{align}

%=============================================================================

\section{SMEFT coefficients in the mass basis}
\label{app:mass-basis}

In the up-mass basis we absorb the unitary rotations $S_{L,R}^u$ between the flavor and mass basis into the Wilson coefficients.
The ones of  the operators \eqref{eq:top_boson}  are then given as
\begin{align}
\begin{aligned}
    \hat C_{\varphi q}^{(1)ij}  &= \hat C_{\varphi q}^{(1)kl} \left(S^{u\dagger}_L\right)_{ik}  \left(S^{\vphantom{\dagger}u}_L\right)_{lj} \,,\enskip 
    &\hat C_{\varphi q}^{(3)ij}  &= \hat C_{\varphi q}^{(3)kl} \left(S^{u\dagger}_L\right)_{ik}  \left(S^{\vphantom{\dagger}u}_L\right)_{lj} \,,\\
    \hat C_{uB}^{ij}  &= \hat C_{uB}^{kl} \left(S^{u\dagger}_L\right)_{ik}  \left(S^{\vphantom{\dagger}u}_R\right)_{lj} \,,\enskip 
    &\hat C_{uW}^{ij}  &= \hat C_{uW}^{kl} \left(S^{u\dagger}_L\right)_{ik}  \left(S^{\vphantom{\dagger}u}_R\right)_{lj} \,,\enskip \\
    \hat C_{uG}^{ij}  &= \hat C_{uG}^{kl} \left(S^{u\dagger}_L\right)_{ik}  \left(S^{\vphantom{\dagger}u}_R\right)_{lj} \,,\enskip
    &\hat C_{\varphi u}^{ij}  &= \hat C_{\varphi u}^{kl} \left(S^{u\dagger}_R\right)_{ik}  \left(S^{\vphantom{\dagger}u}_R\right)_{lj} \,.\enskip 
    \label{eq:mass-boson}
\end{aligned}
\end{align}
Similarly, we obtain for the coefficients of  the four-fermion operators \eqref{eq:top_fermion} 
\begin{align}
    \begin{aligned}
        \hat C^{(1)ij}_{l q} &= C^{(1)kl}_{l q} \left(S^{u\dagger}_L\right)_{ik}  \left(S^{\vphantom{\dagger}u}_L\right)_{lj} \,,\enskip
        &\hat C^{(3)ij}_{l q} &= C^{(3)kl}_{l q} \left(S^{u\dagger}_L\right)_{ik}  \left(S^{\vphantom{\dagger}u}_L\right)_{lj} \,,\\
        \hat C^{ij}_{qe} &= C^{kl}_{qe} \left(S^{u\dagger}_L\right)_{ik}  \left(S^{\vphantom{\dagger}u}_L\right)_{lj} \,,
        &\hat C^{ij}_{eu} &= C^{kl}_{eu} \left(S^{u\dagger}_R\right)_{ik}  \left(S^{\vphantom{\dagger}u}_R\right)_{lj} \,,\enskip\\
        \hat C^{ij}_{l u} &= C^{kl}_{l u} \left(S^{u\dagger}_R\right)_{ik}  \left(S^{\vphantom{\dagger}u}_R\right)_{lj} \,.
        \label{eq:mass-fermion}
    \end{aligned}
\end{align}

\section{SMEFT operators in the mass basis}
\label{app:relations}

In the up-mass eigenbasis, with coefficients defined according to Eq.~\eqref{eq:mass-boson} we find for the effective operators in Eq.~\eqref{eq:top_boson}
\begin{align}
\begin{aligned}
    \hat C_{\varphi q}^{(1)ij}  \hat O_{\varphi q}^{(1)ij} =& \hat C_{\varphi q}^{(1)ij} \left(\varphi^\dagger i\overleftrightarrow D_\mu \varphi\right)\left(\bar u_L^{\prime i} \gamma^\mu  u_L^{\prime j} + V^\dagger_{ki} V^{\vphantom{\dagger}}_{il}  \bar d_L^{\prime k} \gamma^\mu  d_L^{\prime l}\right)\,,\\
    \hat C_{\varphi q}^{(3)ij}  \hat O_{\varphi q}^{(3)ij} =& \hat C_{\varphi q}^{(3)ij} \left(\varphi^\dagger i\overleftrightarrow D_\mu^3 \varphi\right)\left( \bar u_L^{\prime i} \gamma^\mu  u_L^{\prime j} - V^\dagger_{ki} V^{\vphantom{\dagger}}_{jl}  \bar d_L^{\prime k} \gamma^\mu  d_L^{\prime l}\right) + \dots\,,\\
    \hat C_{uB}^{ij} \hat O_{uB}^{ij} =& \hat C_{uB}^{ij} \left(\bar{u}^{\prime i}_L\sigma^{\mu\nu} u_R^{\prime j}\right)\frac{h + v}{\sqrt{2}}B_{\mu\nu} +\textmd{h.c.}\,,\\
    \hat C_{uW}^{ij} \hat O_{uW}^{ij} =& \hat C_{uW}^{ij} \left[\left(\bar{u}^{\prime i}_L\sigma^{\mu\nu} u_R^{\prime j}\right)\frac{h + v}{\sqrt{2}}W^3_{\mu\nu}
    + V^{{\dagger}}_{ki} \left(\bar{d}^{\prime k}_L\sigma^{\mu\nu} u_R^{\prime j}\right)\frac{h + v}{\sqrt{2}}W^-_{\mu\nu}\right]+\textmd{h.c.}\,,\\
    \hat C_{uG}^{ij} \hat O_{uG}^{ij} =& \hat C_{uG}^{ij} \left(\bar{u}^{\prime i}_L\sigma^{\mu\nu}T^{A} u_R^{\prime j}\right)\frac{h + v}{\sqrt{2}}G_{\mu\nu}^{A}+\textmd{h.c.} \,,\\
    \hat C_{\varphi u}^{ij}  \hat O_{\varphi u}^{ij} =&  \hat C_{\varphi u}^{ij} \left(\varphi^\dagger i\overleftrightarrow D_\mu \varphi\right)\left(\bar u_R^i \gamma^\mu u_R^j\right)\,.
    \end{aligned}
\end{align}
Similarly, we find for the four-fermion operators in Eq.~\eqref{eq:top_fermion} with coefficients defined in Eq.~\eqref{eq:mass-fermion} 
\begin{align}
    \begin{aligned}
        \hat C_{lq}^{(1)ij}  \hat O_{lq}^{(1)ij} =& \hat C_{lq}^{(1)ij} \left(\bar l_L \gamma_\mu l_L\right)\left(\bar u_L^{\prime i} \gamma^\mu  u_L^{\prime j} + V^\dagger_{ki} V^{\vphantom{\dagger}}_{il}  \bar d_L^{\prime k} \gamma^\mu  d_L^{\prime l}\right)\,,\\
        \hat C_{lq}^{(3)ij}  \hat O_{lq}^{(3)ij} =& \hat C_{lq}^{(3)ij} \left(\bar l_L \gamma_\mu \tau^3 l_L\right)\left(\bar u_L^{\prime i} \gamma^\mu  u_L^{\prime j} - V^\dagger_{ki} V^{\vphantom{\dagger}}_{il}  \bar d_L^{\prime k} \gamma^\mu  d_L^{\prime l}\right) + \dots\,,\\
        \hat C_{q e}^{(1)ij}  \hat O_{q e}^{(1)ij} =& \hat C_{q e}^{(1)ij} \left(\bar e_R \gamma_\mu e_R\right)\left(\bar u_L^{\prime i} \gamma^\mu  u_L^{\prime j} + V^\dagger_{ki} V^{\vphantom{\dagger}}_{il}  \bar d_L^{\prime k} \gamma^\mu  d_L^{\prime l}\right)\,,\\
        \hat C_{eu}^{ij}  \hat O_{e u}^{ij} =& \hat C_{eu}^{ij} \left(\bar e_R \gamma_\mu e_R\right)\left(\bar u_R^{\prime i} \gamma^\mu  u_R^{\prime j} \right)\,,\\
        \hat C_{lu}^{ij}  \hat O_{lu}^{ij} =& \hat C_{lu}^{ij} \left(\bar l_L \gamma_\mu l_L\right)\left(\bar u_R^{\prime i} \gamma^\mu  u_R^{\prime j} \right)\,.
    \end{aligned}
\end{align}
These results are in agreement with Ref.~\cite{Aebischer:2015fzz}.

\section{Analytic formulas for one-loop matching}
\label{App:match}
The contributions from $\tilde C^{(i)}_{\varphi q}$ to $C_{7}$ and $C_{8}$ are taken from \cite{Dekens:2019ept}:
\begin{align}
    E_7^{\varphi q}(x_t)&=\frac{1}{27}(2\cos^2\theta_w+1)\,,\\
    E_7^{\varphi q(3)}(x_t)&=\frac{8 x_t^3 + 5 x_t^2 - 7 x_t}{12 (x_t - 1)^3} + \frac{x_t^2 (2 - 3 x_t)}{2 (1 - x_t)^4} \log x_t + \frac{1}{54}(-4\cos^2\theta_w+67)\,,\\
    E_8^{\varphi q}(x_t)&=\frac{1}{9}(2\cos^2\theta_w+1)\,,\\
    E_8^{\varphi q(3)}(x_t)&=-\frac{x_t (x_t^2 - 5 x_t - 2)}{4 (x_t - 1)^3} - \frac{3}{2}\frac{x_t^2}{(x_t - 1)^4} \log x_t + \frac{1}{9}(2\cos^2\theta_w+1)\,.
\end{align}
The remaining functions relevant for contributions from dipole operators read \cite{Aebischer:2015fzz}
\begin{align}
    &\begin{aligned}
        E_{7}^{uW}(x_t)&=
        \frac{-9 x_{t}^3+63 x_{t}^2-61 x_{t}+19}{48 \left(x_{t}-1\right)^3}
        +\frac{\left(3 x_{t}^4-12 x_{t}^3-9 x_{t}^2+20 x_{t}-8\right) \ln \left(x_{t}\right)}{24 \left(x_{t}-1\right)^4}\\
        &+\frac{1}{8}\ln\left(\frac{m_W^2}{\mu_W^2}\right)\,,
        \end{aligned}\\
        &F_{7}^{uW}(x_t)=\frac{x_{t} \left(2-3 x_{t}\right) \ln \left(x_{t}\right)}{4 \left(x_{t}-1\right)^4}
        -\frac{3 x_{t}^3-17 x_{t}^2+4 x_{t}+4}{24 \left(x_{t}-1\right)^3}\,,\\
        &E_{7}^{uB}(x_t)=-\frac{1}{8} \ln \left(\frac{m_W^2}{\mu_W^2}\right)-\frac{\left(x_{t}+1\right)^2}{16 \left(x_{t}-1\right)^2}-\frac{x_{t}^2 \left(x_{t}-3\right) \ln \left(x_{t}\right)}{8 \left(x_{t}-1\right)^3}\,,\\
        &F_{7}^{uB}(x_t)=-\frac{1}{8}\,,\\
        &E_8^{uW}(x_t)=\frac{3 x_{t}^2-13 x_{t}+4}{8 \left(x_{t}-1\right)^3}
        +\frac{\left(5 x_{t}-2\right) \ln \left(x_{t}\right)}{4 \left(x_{t}-1\right)^4}\,,\\
        &F_8^{uW}(x_t)=\frac{x_{t}^2-5 x_{t}-2}{8 \left(x_{t}-1\right)^3}+\frac{3 x_{t} \ln \left(x_{t}\right)}{4 \left(x_{t}-1\right)^4}\,,\\
        &E_8^{uG}(x_t)=E_7^{uB}(x_t)\,,\\
        &F_8^{uG}(x_t)=F_7^{uB}(x_t)\,.
\end{align}

The following functions relevant for the matching of up-type dipole operators on $C_9$ and $C_{10}$ are taken from Ref.~\cite{Aebischer:2015fzz} and read
\begin{align}
    Y_{ uW} (x_t) &=  \frac{3x_t}{4(x_t-1)} -\frac{3 x_t}{4(x_t-1)^2}\ln \left(x_t\right)\,,\\
	Z_{ uW}(x_t) &= \frac{99x_t^3-136x_t^2-25x_t+50}{36(x_t-1)^3} -\frac{24x_t^3-45x_t^2+17x_t+2}{6(x_t-1)^4}\ln \left(x_t\right)\,,\\
	Z_{ uB} (x_t) &= -\frac{x_t^2+3x_t-2}{4(x_t-1)^2}  +\frac{3x_t-2}{2(x_t-1)^3}\ln \left(x_t\right)\,.
\end{align}
Contributions for both four-fermion operators and operators with two Higgs bosons can be parametrized in terms of functions \cite{Endo:2020kie}
\begin{align}
    K_0(x,\mu) &= 
    - \frac{x}{32} \left[
    \ln\frac{\mu^2}{m_W^2}
    + \frac{3(x+1)}{2(x-1)} 
    - \frac{x^2-2x+4}{(x-1)^2}\ln x
    \right]\,,
    \\
    K_1(x,\mu) &= 
    \frac{x}{16} \left[
    \ln\frac{\mu^2}{m_W^2} 
    + \frac{x-7}{2(x-1)} 
    - \frac{x^2-2x-2}{(x-1)^2} \ln x
    \right]\,,
    \\
    K_2(x,\mu) &= 
    - \frac{x}{8} \left[
    \ln\frac{\mu^2}{m_W^2} 
    + 1 - \ln x
    \right]\,,\\
    J_2(x) &= \frac{x}{8}\,,\\
    J_3(x,\mu) &= 
    -\frac{3}{16}x_t \left[ 
    \ln\frac{\mu^2}{m_W^2}
    + \frac{x_t+3}{2(x_t-1)} 
    - \frac{x_t^2+1}{(x_t-1)^2} \ln x_t 
    \right]\,,\\
    B(x) &= 
    \frac{3}{16}x \left[ 
      \frac{1}{x-1} 
    - \frac{1}{(x-1)^2} \ln x 
    \right]\,,\\
    D(x) &= 
    -\frac{2}{9} \ln x
    -\frac{x}{72} \left[ 
    \frac{82x^2-151x+63}{(x-1)^3} 
    - \frac{10x^3+59x^2-138x+63}{(x-1)^4} \ln x 
    \right]\,.
\end{align}
With these definitions, the functions appearing in the matching to $C_9$ and $C_{10}$ read \cite{Endo:2020kie}
\begin{align}
    &\begin{aligned}
        I_1(x_t)&= -J_2(x_t) - 2 K_0(x_t,\mu_W) \\
        &= \frac{x_t}{16} \left[ 
            \ln\frac{\mu^2}{m_W^2}
          - \frac{x_t-7}{2(x_t-1)} 
          - \frac{x_t^2-2x_t+4}{(x_t-1)^2} \ln x_t 
          \right].
    \end{aligned}\\
    &\begin{aligned}
        I_2(x_t)=& J_2(x_t) + K_2(x_t,\mu_w) + K_0(x_t,\mu)\\
        =&-\frac{x_t}{32}\left(5 \ln\frac{\mu_W^2}{m_W^2} + \frac{3(x_t+1)}{2(x_t-1)} - \frac{5x_t^2 -10x_t + 8}{(x_t-1)^2} \ln x_t\right)\,,
    \end{aligned}\\
    &\begin{aligned}
        I^{lq}(x_t) =& -J_2(x_t) + K_2(x_t,\mu_w) + K_0(x_t,\mu)\\
        =&-\frac{x_t}{32}\left( 8 + 5 \ln\frac{\mu_W^2}{m_W^2} + \frac{3(x_t+1)}{2(x_t-1)} - \frac{5x_t^2 -10x_t + 8}{(x_t-1)^2} \ln x_t\right)\,,
    \end{aligned}\\
    &\begin{aligned}
        I_1^{\varphi q}(x_t) =& (-1+4\sin^2\theta_w) \left(J_2(x_t) + 2 J_3(x_t) - K_1(x_t) - 3 K_0(x_t,\mu)\right)\\ &+ 2\left( B(x_t) + 2 \sin^2\theta_w D(x_t)\right)\\
        =&(-1+4\sin^2\theta_w)\frac{x_t}{32}\left( 4 -11 \ln\frac{\mu_W^2}{m_W^2} - \frac{5x_t+13}{2(x_t-1)} + \frac{11x_t^2 +2x_t - 4}{(x_t-1)^2} \ln x_t\right)\\
        &+    \frac{3}{8}x \left[ \frac{1}{x_t-1} - \frac{1}{(x_t-1)^2} \ln x_t \right] + \sin^2\theta_w   \left\{  -\frac{8}{9} \ln x_t \right.\\
        &- \left.\frac{x_t}{18} \left[ \frac{82x_t^2-151x_t+63}{(x_t-1)^3} 
        - \frac{10x_t^3+59x_t^2-138x_t+63}{(x_t-1)^4} \ln x_t \right] \right\}\,,
    \end{aligned}\\
    &\begin{aligned}
        I_2^{\varphi q}(x_t) =&\left(J_2(x_t) + 2 J_3(x_t) - K_1(x_t) - 3 K_0(x_t,\mu)\right)+ 2 B(x_t)\\
        =&\frac{x_t}{32}\left( 4 -11 \ln\frac{\mu_W^2}{m_W^2} - \frac{5x_t+37}{2(x_t-1)} + \frac{11x_t^2 +2x_t + 8}{(x_t-1)^2} \ln x_t\right)\,,
    \end{aligned}
\end{align}
where we neglected CKM-suppressed contributions $\sim |V_{ts}|^2,\sim |V_{td}|^2$, which are smaller by a factor of at least $\sim 10^{-3}$.

The functions $H_i$ relevant for the matching of $\tilde{C}^{(1)}_{\varphi q}$ and $\tilde{C}^{(3)}_{\varphi q}$ onto $C_1^\textmd{mix}$ read \cite{Bobeth:2017xry}
\begin{align}
    H_1(x_t) &= -\frac{x_t-7}{4(x_t-1)} - \frac{x_t^2-2x_t+4}{2(x_t-1)^2}\ln x_t\,,\\
    H_1(x_t) &= +\frac{7x_t-25}{4(x_t-1)} - \frac{x_t^2-14x_t+4}{2(x_t-1)^2}\ln x_t\,.
\end{align}

Finally, functions relevant for the matching of SMEFT coefficients onto $C_L$ at one-loop level are taken from \cite{Dekens:2019ept}. 
Here, we give results with all evanescent coefficients set to 1:
\begin{align}
    &\begin{aligned}
        I^{\nu }_{uW} &= \frac{m_t}{m_W\sin^2\theta_w}\left(-\frac{3(x_t-2)}{4\sqrt 2(x_t-1)} - \frac{3x_t\ln x_t}{\sqrt 2(x_t-1)^2}\right)\,,
    \end{aligned}\\
    &\begin{aligned}
        I^{\nu (1)}_{\varphi q} &= \frac{1}{\sin^2\theta_w}\left(\frac{x_t}{8} - \frac{3 x_t(x_t+1)}{32(x_t-1)}-\frac{x_t(x_t^2-2x_t+4)\ln\frac{\mu_W^2}{m_t^2}}{16(x_t-1)^2} + \frac{3 x_t\ln\frac{\mu_W^2}{m_W^2}}{16\left(x_t-1\right)^2}\right.\\
        &+\left. \frac{(2m_W^2+m_Z^2)}{8m_W^2}-\frac{3x_t\ln\frac{\mu_W^2}{m_t^2}}{8}\right)\,,
    \end{aligned}\\
    &\begin{aligned}
        I^{\nu (3)}_{\varphi q} &= \frac{1}{\sin^2\theta_w}\left(-\frac{x_t}{8} + \frac{5 x_t(x_t-7)}{32(x_t-1)}+\frac{x_t(7x_t^2-2x_t-20)\ln\frac{\mu_W^2}{m_t^2}}{16(x_t-1)^2} - \frac{3 x_t(4x_t-9)\ln\frac{\mu_W^2}{m_W^2}}{16\left(x_t-1\right)^2}\right.\\ 
        &+\left. \frac{19m_W^2+m_Z^2}{8m_W^2}-\frac{3\ln\frac{mu_W^2}{m_W^2}}{8}+\frac{3m_W^2\ln\frac{\mu_W^2}{m_W^2}}{4m_W}\right)\,,
    \end{aligned}\\
    &\begin{aligned}
        I^{\nu }_{l u} &= \frac{1}{\sin^2\theta_w}\left(-\frac{x_t(x_t-7)}{32(x_t-1)} + \frac{(x_t^3-2x_t^2+4x_t)\ln\frac{\mu_w^2}{m_t^2}}{16(x_t-1)^2}-\frac{3x_t\ln\frac{\mu_w^2}{m_W^2}}{16(x_t-1)^2}\right)\,,
    \end{aligned}\\
    &\begin{aligned}
        I^{\nu (1)}_{l q} &= \frac{1}{\sin^2\theta_w}\left(\frac{x_t}{8} - \frac{3 x_t(x_t+1)}{32(x_t-1)}-\frac{x_t(x_t^2-2x_t+4)\ln\frac{\mu_W^2}{m_t^2}}{16(x_t-1)^2} + \frac{3 x_t\ln\frac{\mu_W^2}{m_W^2}}{16\left(x_t-1\right)^2}\right. \\ 
        &+\left. \frac{11(2m_W^2+m_Z^2)}{48m_W^2}+\frac{(2m_W^2+m_Z^2)\ln\frac{\mu_W^2}{m_Z^2}}{8m_W}\right)\,,
    \end{aligned}\\
    &\begin{aligned}
        I^{\nu (3)}_{l q} &= \frac{1}{\sin^2\theta_w}\left(\frac{x_t}{8} - \frac{3 x_t(x_t+1)}{32(x_t-1)}-\frac{x_t(x_t^2-26x_t+28)\ln\frac{\mu_W^2}{m_t^2}}{16(x_t-1)^2} + \frac{3 x_t(8x_t-9)\ln\frac{\mu_W^2}{m_W^2}}{16\left(x_t-1\right)^2} \right.\\ 
        &+ \left. \frac{-154m_W^2-11m_Z^2}{48m_W^2}-\frac{3\ln\frac{mu_W^2}{m_W^2}}{2}-\frac{(2m_W^2+m_Z^2)\ln\frac{\mu_W^2}{m_Z^2}}{8m_W}\right)\,.
    \end{aligned}
\end{align}

\section{Numerical matching conditions}
\label{app:numerics}

The numerical values of the tree-level matching conditions in Eq.~\eqref{eq:match-tree} read at $\mu_W=m_W$
\begin{align}
    \begin{aligned}
        \Delta C_9^\textmd{tree} &= 402.1 \left[ \tilde C^{1}_{lq} + \tilde C^{3}_{lq} + \tilde C_{q e}\right] -44.53 \left(\tilde C^{1}_{\varphi q} + \tilde C^{3}_{\varphi q}\right)\,,\\
        \Delta C_{10}^\textmd{tree} &= 402.1 \left[- \tilde C^{1}_{lq} - \tilde C^{3}_{lq} + \tilde C_{q e} + \tilde C^{1}_{\varphi q} + \tilde C^{3}_{\varphi q} \right]\,,\\
        \Delta C_{L}^\textmd{tree} &= 402.1 \left[ \tilde C^{1}_{lq} - \tilde C^{3}_{lq} + \tilde C^{1}_{\varphi q} + \tilde C^{3}_{\varphi q}\right]\,.
        \label{Eq:Matchtree-numerics}
    \end{aligned}
\end{align}
For the one-loop contributions in Eqs.~\eqref{Eq:MatchC7}-\eqref{Eq:MatchC1} we obtain at $\mu_W=m_W$
\begin{align}
    &\begin{aligned}
        \Delta C_{7}^\textmd{loop}=&-2.31\tilde C_{uB} +  0.0925 \tilde C_{uW} - -0.132 \tilde C_{\varphi q}^{(1)} + 1.12 \tilde C_{\varphi q}^{(3)}\,,    
        \label{Eq:MatchC7-numerics}
    \end{aligned}\\
    &\begin{aligned}
        \Delta C_{8}^\textmd{loop}=&-0.669 \tilde C_{uG} +0.271 \tilde C_{uW} + 0.392 \tilde C^{(1)}_{\varphi q} + 1.05 \tilde C^{(3)}_{\varphi q} \,,    
        \label{Eq:MatchC8-numerics}
    \end{aligned}\\
    &\begin{aligned}
        \Delta C_{9}^\textmd{loop}= &2.170\tilde C_{uW} + 2.512 \tilde C_{uB} - 1.81 \tilde C_{\varphi q}^{(1)}  - 1.96 \tilde C_{\varphi q}^{(3)} + 0.148 \tilde C_{\varphi u}\\&- 1.898 \left(\tilde C_{eu}+\tilde C_{lu} \right) -2.242\left(\tilde C^{(1)}_{lq}-\tilde C_{qe}\right)
        -4.444\tilde C^{(3)}_{lq}\,,
        \label{Eq:MatchC9-numerics}
    \end{aligned}\\
    &\begin{aligned}
        \Delta C_{10}^\textmd{loop}=&-7.54 \tilde C_{uW} + 12.8 \tilde C_{\varphi q}^{(1)} - 3.43 \tilde C_{\varphi q}^{(3)} - 1.90 \tilde C_{\varphi u} - 1.90 \left(\tilde C_{eu}-\tilde C_{lu}\right) \\ &- 2.242\left(\tilde C^{(1)}_{lq}- \tilde C_{qe}\right)
        +4.444\tilde C^{(3)}_{lq}\,,
        \label{Eq:MatchC10-numerics}
    \end{aligned}\\
    &\begin{aligned}
        \Delta C_{L}^\textmd{loop}=&-2.88\tilde C_{uW} + 14.9 \tilde C_{\varphi q}^{(1)} + 0.332 \tilde C_{\varphi q}^{(3)} - 1.90 \left(\tilde C_{\varphi u} +\tilde C_{lu}\right)\\&+ 4.622 \tilde C^{(1)}_{lq}+1.033\tilde C^{(3)}_{lq}\,,
        \label{Eq:MatchCL-numerics}
    \end{aligned}\\
    &\begin{aligned}
        \Delta C_{1,tt}^\textmd{mix, loop}=& 4.12\tilde C_{uW} + 14.8 \tilde C_{\varphi q}^{(1)}  + 11.6 \tilde C_{\varphi q}^{(3)}  \,.
        \label{Eq:MatchC1-numerics}
    \end{aligned}
\end{align}

\section{Auxiliary Plots }
\label{app:far}
\begin{figure}[h]
    \centering
    \includegraphics[width=\textwidth]{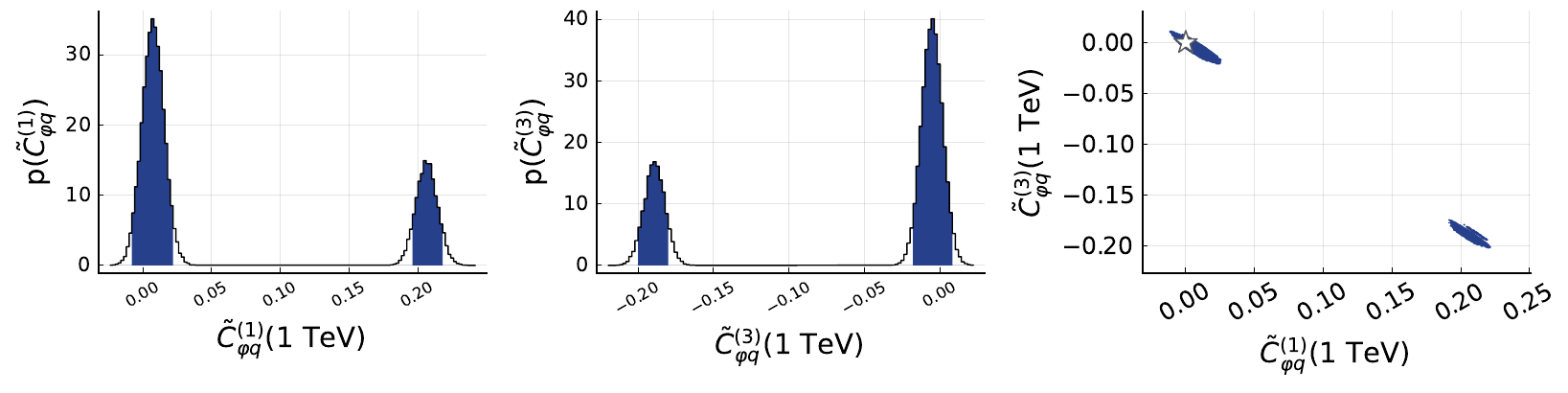}
    \caption{One-dimensional (left, middle) and two-dimensional (right) projections of the posterior distribution for $\tilde C^{(1)}_{\varphi q}$ and $\tilde C^{(3)}_{\varphi q}$. 
    Results are obtained for a fit of eight SMEFT coefficients in Eq.~\eqref{eq:max-dof} to the combined set of present top-quark, $Zb\bar b$, and $B$ physics data.
    The star in the rightmost plot denotes the SM point.
    }
    \label{fig:combined-now-1D-Deviations}
\end{figure}
\begin{figure}[h]
    \centering
    \includegraphics[width=\textwidth]{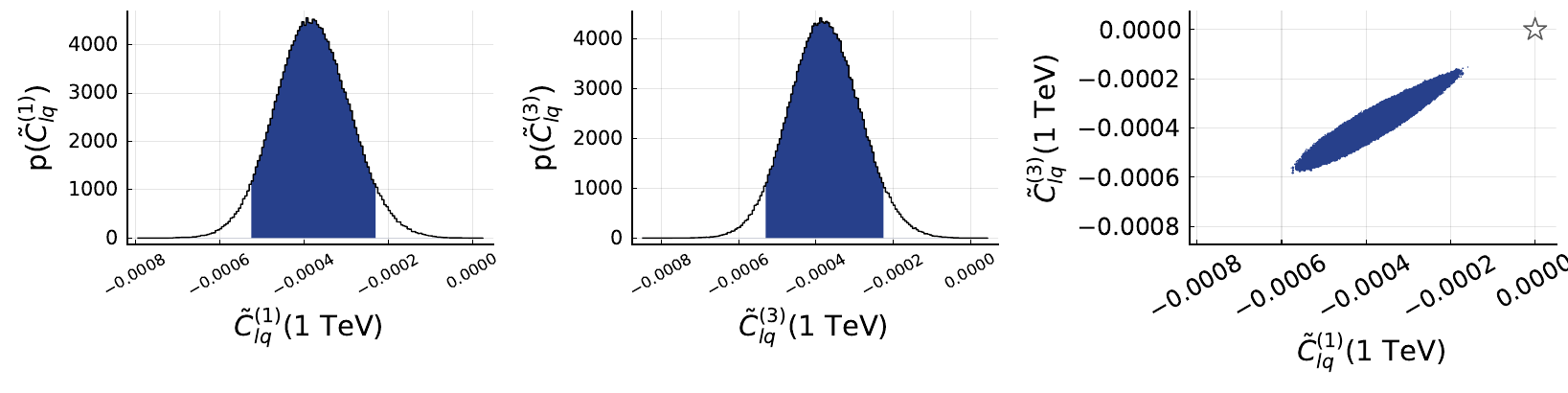}
    \caption{One-dimensional (left, middle) and two-dimensional (right) projections of the posterior distribution for $\tilde C^{(1)}_{lq}$ and $\tilde C^{(3)}_{lq}$. 
    Results are obtained for a fit of all eleven coefficients in Eq.~\eqref{eq:future-SMEFT-coff} to the combined set of present data, near future projections, and CLIC projections for top-quark observables.
    The star in the rightmost plot denotes the SM point. 
    }
    \label{fig:combined-far-1D-Deviations}
\end{figure}
\begin{figure}[h]
    \centering
    \includegraphics[width=\textwidth]{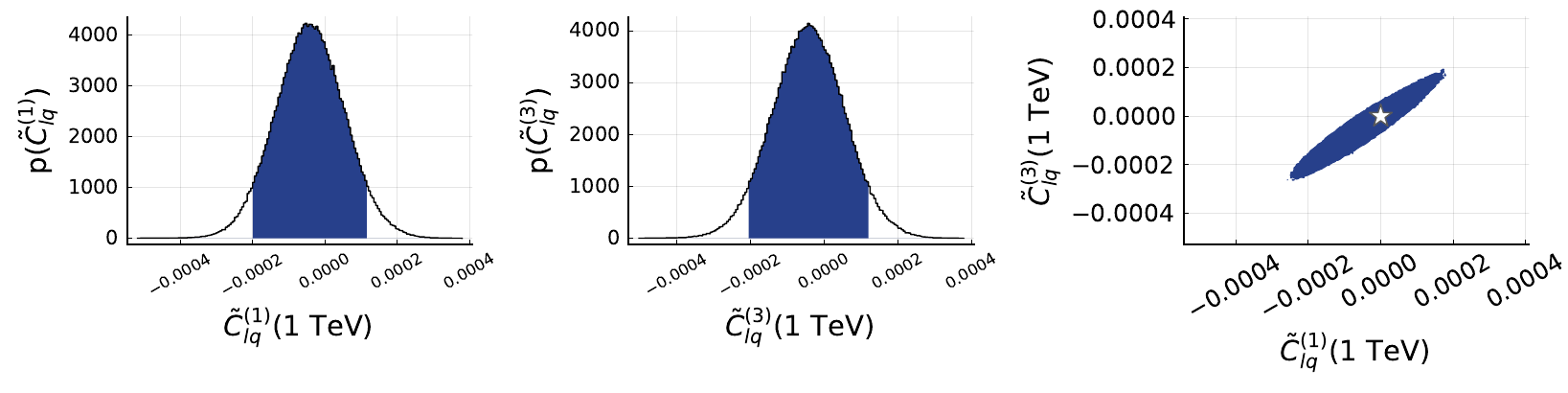}
    \caption{The same as  in Fig.~\ref{fig:combined-far-1D-Deviations} but assuming  SM central values for $b \to s$ observables at Belle II.  }
        \label{fig:combined-far-1D-noDeviations}
\end{figure}

\bibliography{references}
\end{document}